\documentclass[preprint2,numberedappendix,tighten,twocolappendix]{aastex6}  

\usepackage{mathtools,hyperref}
\usepackage{amsmath}
\usepackage{graphicx}
\usepackage[figuresright]{rotating}
\usepackage{natbib}
\usepackage[all]{hypcap}								
\usepackage{braket}
\usepackage{fancyhdr}
\usepackage{bm}
\def\bA{\mathbf{A}}
\def\mA{\mathrm{A}}
\def\bB{\mathbf{B}}

\def\bw{\mathbf{w}}
\def\mw{\mathrm{w}}

\def\bd{\mathbf{d}}
\def\md{\mathrm{d}}

\def\bk{\mathbf{k}}

\def\bb{\mathbf{b}}

\def\bR{\mathbf{R}}
\def\mR{\mathrm{R}}

\def\bC{\mathbf{C}}

\def\bx{\mathbf{x}}

\def\by{\mathbf{y}}

\def\mcL{\mathcal{L}}
\def\bL{\mathbf{L}}
\def\mK{\mathrm{K}}
\def\mk{\mathrm{k}}
\def\bK{\mathbf{K}}
\def\bI{\mathbf{I}}

\def\mE{\mathrm{E}}
\def\mS{\mathrm{S}}

\def\bSigma{\boldsymbol{\Sigma}}

\def\btheta{\boldsymbol{\theta}}

\def\bPhi{\boldsymbol{\Phi}}
\def\bgamma{\boldsymbol{\gamma}}

\def\bLambda{\boldsymbol{\Lambda}}

\def\bdelta{\boldsymbol{\delta}}
\def\tocm{$21\,\textrm{cm}$\ }

\def\Tb{\delta T_{\text{b}}}
\def\Rmfp{R_{\text{mfp}}}

\def\Tvirmin{T_{\text{vir}}^{\text{min}}}
\def\fX{f_{X}}
\def\aX{\alpha_{X}}
\def\numin{\nu_{\text{min}}}
\def\xHI{x_{\text{HI}}}

\def\Tg{T_{\gamma}}
\def\Ts{T_{\text{S}}}
\def\se{\sigma_{8}}
\def\Hn{H_{0}}
\def\Ombhh{\Omega_{b}h^{2}}
\def\Omchh{\Omega_{c}h^{2}}
\def\ns{n_{s}}

\def\fesc{f_{\text{esc}}}

\def\fcoll{f_{\text{coll}}}

\def\DD{\Delta^{2}_{21}}
\def\kpara{k_{\parallel}}
\def\kperp{k_{\perp}}

\begin{document}
\title{Emulating Simulations of Cosmic Dawn for \tocm Power Spectrum Constraints on Cosmology, Reionization, and X-ray Heating} 
\shorttitle{Emulating Simulations of Cosmic Dawn for \tocm Parameter Constraints}
\author{
Nicholas S. Kern\altaffilmark{1}$^{,\star}$,
Adrian Liu\altaffilmark{1}$^{,\dagger}$,
Aaron R. Parsons\altaffilmark{1},
Andrei Mesinger\altaffilmark{2},
Bradley Greig\altaffilmark{2}
}
\shortauthors{N. Kern et al.}

\altaffiltext{1}{Department of Astronomy and Radio Astronomy Laboratory, University of California Berkeley, Berkeley, CA 94720, USA}
\altaffiltext{2}{Scuola Normale Superiore, Piazza dei Cavalieri 7, I-56126 Pisa, Italy}

\newcommand{\nsk}[1]{{\color{blue} \textbf{[NSK:  #1]}}}
\newcommand{\acl}[1]{{\color{red} \textbf{[ACL:  #1]}}}
\newcommand{\arp}[1]{{\color{cyan} \textbf{[ARP:  #1]}}}
\newcommand{\am}[1]{{\color{violet} \textbf{[AM:  #1]}}}
\newcommand{\bg}[1]{{\color{orange} \textbf{[BG:  #1]}}}

\begin{abstract}
Current and upcoming radio interferometric experiments are aiming to make a statistical characterization of the high-redshift \tocm fluctuation signal spanning the hydrogen reionization and X-ray heating epochs of the universe. However, connecting \tocm statistics to underlying physical parameters is complicated by the theoretical challenge of modeling the relevant physics at computational speeds quick enough to enable exploration of the high dimensional and weakly constrained parameter space. In this work, we use machine learning algorithms to build a fast emulator that can accurately mimic an expensive simulation of the \tocm signal across a wide parameter space. We embed our emulator within a Markov-Chain Monte Carlo framework in order to perform Bayesian parameter constraints over a large number of model parameters, including those that govern the Epoch of Reionization, the Epoch of X-ray Heating, and cosmology. As a worked example, we use our emulator to present an updated parameter constraint forecast for the Hydrogen Epoch of Reionization Array experiment, showing that its characterization of a fiducial \tocm power spectrum will considerably narrow the allowed parameter space of reionization and heating parameters, and could help strengthen \emph{Planck}'s constraints on $\se$. We provide both our generalized emulator code and its implementation specifically for \tocm parameter constraints as publicly available software.
\end{abstract}
\maketitle

\section{Introduction}
\label{sec:intro}	

{\let\thefootnote\relax\footnote{$^{\star}$\href{mailto:nkern@berkeley.edu}{nkern@berkeley.edu}}}
{\let\thefootnote\relax\footnote{$^{\dagger}$Hubble Fellow}}
\setcounter{footnote}{0}

Cosmic Dawn is a fundamental milestone in our universe's history, and marks the era when the first generation of stars and galaxies formed, ending the Dark Ages that followed recombination. These first luminous sources eventually reionized the neutral hydrogen filling the Intergalactic Medium (IGM) during the Epoch of Reionization (EoR). For all of their implications on the formation and evolution of the first galaxies and compact objects, the EoR and Cosmic Dawn remain a relatively unexplored portion of our universe's history. However, in recent years there have been significant observational advances in our understanding of this epoch. These include Cosmic Microwave Background (CMB) measurements that constrain the timing of reionization \citep{Planck2016, Zahn2012, Mesinger2012}; direct measurements of the bright end of the ultraviolet luminosity function of galaxies up to $z \sim 10$, which constrain some of the sources of reionization \citep{Bouwens2015, Finkelstein2015, Livermore2017}, and Lyman-$\alpha$ absorption studies that place limits on the end of reionization \citep{Fan2006,Becker2015,McGreer2015}. 

Another promising class of probes are radio interferometer intensity mapping experiments targeting the \tocm hyperfine transition from neutral hydrogen \citep{Hogan1979,Scott1990,Madau1997,Tozzi2000}. Such experiments aim to tomographically map out the distribution, thermal state and ionization state of neutral hydrogen in the IGM throughout Cosmic Dawn, and are potentially the only \emph{direct} probes of the epochs relevant to the formation of the first generations of stars, galaxies, stellar-mass black holes, supernovae, and quasars. For reviews of \tocm cosmology, see e.g. \citet{Furlanetto2006a,Morales2010,Pritchard2012,Loeb2013,Mesinger2016b}. While \tocm cosmology faces formidable observational challenges, recent years have seen significant advances toward resolving issues of optimal array design \citep{Beardsley2012,Parsons2012a,Greig2015b,Dillon2016}, internal systematics \citep{Ewall-Wice2016b,Barry2016,Patil2016,Ewall-Wice2016c}, and astrophysical foreground mitigation \citep{Datta2010,Morales2012,Vedantham2012,Parsons2012a,Trott2012,Chapman2012,Chapman2013,Thyagarajan2013,Pober2013b,Liu2014a,Liu2014b,Switzer2014,Wolz2014,Moore2015,Thyagarajan2015b,Thyagarajan2015a,Asad2015,Chapman2016,Pober2016,Kohn2016,Liu2016b}. Increasingly competitive upper limits have been placed on the redshifted \tocm signal, using instruments such as the Donald C. Backer Precision Array for Probing the Epoch of Reionization (PAPER; \citealt{Parsons2014, Jacobs2015, Ali2015}), the Giant Metrewave Radio Telescope (GMRT; \citealt{Paciga2013}), the Murchison Widefield Array (MWA; \citealt{Dillon2014, Dillon2015b, Ewall-Wice2016b,Beardsley2016}), and the Low Frequency Array (LOFAR; \citealt{Vedantham2015,Patil2017}). Many of these upper limits are stringent enough to be scientifically interesting, and have typically ruled out extremely cold reionization scenarios \citep{Parsons2014,Pober2015,Greig2016}. As these experiments continue to be expanded and second-generation experiments, such as the Hydrogen Epoch of Reionization Array\footnote{\url{http://reionization.org/}} (HERA; \citealt{DeBoer2017}) and the Square Kilometer Array \citep[SKA;][]{Koopmans2015}, begin commissioning and data processing, a first positive detection of the cosmological \tocm signal will soon be within reach.

Following a first detection, instruments such as HERA are expected to make high signal-to-noise measurements of the spatial power spectrum of \tocm brightness temperature fluctuations. Previous studies have shown that such measurements would place stringent constraints on parameters governing the EoR and Epoch of X-ray Heating (EoH) \citep{Pober2014,Liu2016b,Ewall-Wice2016a}, as well as on fundamental cosmological parameters when jointly fit with \emph{Planck} data \citep{McQuinn2006, Mao2008,Barger2009,Clesse2012,Liu2016a}. However, most of these forecasting studies have been limited in at least one of two ways: they have either bypassed full parameter space explorations by employing the Fisher Matrix formalism, or they have relied on simplified parameterizations of the \tocm signal that may not be appropriate for describing real observations. Thus far, the only method capable of systematically exploring the EoR parameter space is \texttt{21CMMC} \citep{Greig2015a}, which combined an optimized version of the semi-numerical simulation \texttt{21cmFAST} \citep{Mesinger2011} with an MCMC sampler. This was used to connect upper limits from PAPER to theoretical models \citep{Greig2016} and to synthesize constraints set by complementary EoR probes \citep{Greig2017a}. However, these studies were limited to $z<10$, because at higher redshifts the inhomogenous heating of the IGM by X-rays becomes important \citep{Kuhlen2006,Pritchard2007,Warszawski2009,Mesinger2013,Pacucci2014,Fialkov2014a,Fialkov2014b,Fialkov2014c,Ghara2015}, and computing it slows down the simulation runtime considerably. As a quantitative illustration, consider \texttt{21cmFAST}, which takes $\sim24$ hours to run on a single core when computing IGM heating. A parameter constraint analysis with 100 MCMC-chains each evaluated for $10^{3}$ steps would take 3 years to run on a 100-core computing cluster, rendering it intractable. This is a problem that must be solved in order for \tocm measurements to place rigorous constraints on theoretical models.

One solution is to optimize the simulations to make them run faster. This was in fact recently accomplished for \texttt{21cmFAST} by \citet{Greig2017b}, who were able to MCMC \texttt{21cmFAST} over EoR and EoH parameters; however, with the inclusion of cosmological parameters this is pushed out of the realm of feasibility. Furthermore, even with detailed optimization, more sophisticated numerical simulations are unlikely to be feasible for MCMC in the near future. Faced with this daunting challenge, one approach is to abandon MCMC parameter fitting altogether. This was explored recently by \citet{Shimabukuro2017a}, who showed that promising results could be obtained using artificial neural networks. If one desires detailed information on constraint uncertainties and parameter degeneracies, however, one must turn to an MCMC framework.

Another solution to the aforementioned problem is to use machine learning algorithms to build surrogate models for the behavior of the expensive simulation. The collection of surrogate models, called an emulator, mimics the simulation across the space of its input parameters. After training the emulator over a pre-computed training set, one can discard the simulation entirely and use the emulator in the MCMC sampler to produce parameter constraints. The speed of the emulator depends on the complexity of the surrogate models, but it is generally many orders of magnitude faster to evaluate than the original simulation. This technique is known as emulation, and has recently taken hold in the astrophysics literature to produce parameter constraints with expensive simulations. Examples within astrophysics include emulation of N-body simulations of the matter power spectrum \citep{Heitmann2006, Habib2007, Heitmann2009, Schneider2011}, simulations of the Cosmic Microwave Background power spectrum \citep{Fendt2007, Aslanyan2015}, simulations of weak lensing \citep{Petri2015}, stellar spectra libraries \citep{Czekala2015}, and numerical relativity gravitational waveforms \citep{Field2014}. In short, emulators allow us to produce parameter constraints with simulations that are otherwise unusable for such purposes. Another crucial benefit of emulators is their repeatability: once we have put in the computational resources and time to build the training set, if we change our measurement covariance matrix or add more data to our observations, re-running the MCMC chains with an emulator for updated fits is extremely quick. Even for semi-numerical simulations that are brought into the realm of MCMC-feasibility via optimization techniques, having to repeat an MCMC analysis many times may be computationally prohibitive. 

In preparation for observations from upcoming \tocm experiments, we have built a fast and accurate emulator for simulations of Cosmic Dawn. Embedding it within an MCMC framework, we present updated forecasts on the constraints that a \tocm power spectrum experiment like HERA will place on EoR \& EoH astrophysical parameters and now also include $\Lambda$CDM base cosmological parameters. It is important to note that the emulator algorithm we present here is not tied to any specific model of Cosmic Dawn. Although we will proceed using a particular simulation of Cosmic Dawn, we could in principle repeat these calculations using an entire suite of various simulations with only minor changes to our procedure. We provide a generalized implementation of our emulator algorithm in a publicly-available Python package called \texttt{emupy}.\footnote{\url{https://github.com/nkern/emupy}} This base package can be used to emulate any dataset and is not specific to \tocm cosmology. We also provide our implementation of the emulator code specific to \tocm---including our \tocm likelihood function, sensitivity forecasts and simulation training sets---in a publicly-available Python package called \texttt{pycape}.\footnote{\url{https://github.com/nkern/pycape}}

The rest of this paper is organized as follows. In \autoref{sec:emu_alg} we provide a detailed overview of our emulator algorithm. In \autoref{sec:simulations} we discuss our Cosmic Dawn simulation and its model parameterization. In \autoref{sec:forecast} we discuss observational systematics for the upcoming HERA experiment and forecast the ability of HERA to constrain astrophysical and cosmological parameters, and in \autoref{sec:discussion} we provide performance benchmarks for further validation of the emulator algorithm. We summarize our conclusions in \autoref{sec:conclusion}.


\section{Building the Emulator}
\label{sec:emu_alg}
At the most basic level, emulation is a combination of three major steps: (i) building a training set, (ii) regressing for analytic functions that mimic the training set data and (iii) evaluating those functions at desired interpolation points and accounting for interpolation error. The emulator itself is then just the collection of these functions, which describe the overall behavior of our simulation. To produce parameter constraints, we simply substitute the simulation with the emulator in our likelihood function, attach it to our MCMC sampler, and let the sampler explore the posterior distribution across our model parameter space. In the following sections, we describe the various steps that go into building such an emulator, which allows us to produce parameter constraints using simulations that would otherwise be either too computationally expensive or take too long to run iteratively.

\subsection{Training Set Design}
\label{sec:exp_design}
To emulate the behavior of a simulation, we first require a training set of simulation outputs spanning our $N$ dimensional parameter space, with each sample corresponding to a unique choice of parameter values $\btheta = \{\theta_{1},\theta_{2},\dots,\theta_{N}\}$, where each $\theta_{i}$ is a vector containing the selected values for our tunable model parameters of our simulation. These, for example, could be cosmological parameters like $\sigma_{8}$ or $H_{0}$. Deciding where in our model parameter space to build up our finite number of training samples is called training set ``design." The goal in creating a particular training set design is to maximize the emulator's accuracy across the model parameter space, while minimizing the number of samples we need to generate. This is particularly crucial for computationally expensive simulations because the construction of the training set will be the most dominant source of overhead. Promising designs include variants of the Latin-Hypercube (LH) design, which seeks to produce uniform sampling densities when all points are marginalized onto any one dimension \citep{McKay1979}. Previous studies applying emulators to astrophysical contexts have shown LH designs to work particularly well for Gaussian-Process based emulators \citep{Heitmann2009}. To generate our LH designs, we use the publicly-available Python software \texttt{pyDOE}\footnote{\url{https://pythonhosted.org/pyDOE}}.

Of particular concern in training set design is the ``curse of dimensionality", or the fact that a parameter space volume depends exponentially on its dimensionality. In other words, in order to sample a parameter space to constant density, the number of samples we need to generate depends exponentially on the dimensionality of the space. One way around this is to impose a spherical prior on our parameter space. This allows us to ignore sampling in the corners of the hypervolume where the prior distribution has very small probability. In low dimensional spaces, this form of cutting corners only marginally helps us; in two dimensions, for example, the area of a square is only $4/\pi$ greater than the area of its circumscribed circle. In ten dimensions, however, the volume of a hypercube is 400 times that of its circumscribed hypersphere. In eleven dimensions this increases to over a factor of 1000. This means that if we choose to restrict ourselves to a hypersphere instead of a hypercube in an eleven dimensional space, we have reduced the volume our training set needs to cover by over three orders of magnitude. \citet{Schneider2011} investigated the benefits of this technique, and used the Fisher Matrix formalism to inform the size of the hypersphere, which they call Latin-Hypercube Sampling Fisher Sphere (LHSFS). This technique works well in the limit that we already have relatively good prior distributions on our parameters. For parameters that are weakly constrained, we may need to turn to other mechanisms for narrowing the parameter space before training set construction.

The parameter constraint forecast we present in \autoref{sec:param_explore}, for example, starts with a coarse rectangular LH design spanning a wide range in parameter values. We emulate at a highly approximate level and use the MCMC sampler to roughly locate the region of high probability in parameter space. We supplement this initial training set with more densely-packed, spherical training sets in order to further refine our estimate of the maximum a posteriori (MAP) point (\autoref{sec:constraints}). The extent of the supplementary spherical training sets are informed from a Fisher Matrix forecast, similar to \citep{Schneider2011}.

Our training sets contain on the order of thousands to tens of thousands of samples. This is not necessitated by our emulator methodology, but by our science goal at hand: our limited empirical knowledge of Cosmic Dawn and EoR means that a dominant source of uncertainty on the \tocm power spectrum comes not from the accuracy of our simulations, but by the allowed range of the many model parameters that affect the power spectrum \citep{Mesinger2013}. As discussed, the more model parameters one incorporates the larger the parameter space volume becomes, and therefore more training samples are typically needed to cover the space. Previous studies applying emulators to the problem of parameter estimation have found success using large training sets to handle high dimensionality \citep{Gramacy2015}. In order to generate large training sets, we are limited to using simulations that are themselves only moderately expensive to run so that we can build up a large training set in a reasonable amount of time. This is our motivation for emulating a semi-numerical simulation of Cosmic Dawn like \texttt{21cmFAST}, which is itself much cheaper to run than a full numerical simulation. We discuss specifics of our adopted model further in \autoref{sec:simulations}.

\subsection{Data Compression}
\label{sec:data_compression}
After constructing a training set, our next task is to decide on which simulation data products to emulate over the high dimensional parameter space. Let us define each number that our simulation outputs as a single datum called $\md$, which in our case will be the \tocm power spectrum, $\DD$, at a specific $k$ mode and a specific redshift $z$.\footnote{See \autoref{eqn:DD} for a formal definition of $\DD$.} Because the power spectra are non-negative quantities, we will hereafter work with the log-transformed data. For example, we might choose our first data element as $\md_{1} = \ln\DD(k=0.1 $ h Mpc$^{-1}, z=10.0).$ We then take all $n$ simulation outputs we would like to emulate and concatenate them into a single column vector,
\begin{align}
\label{eq:ddef}
\bd &= \left(\begin{array}{c}
\md_{1} \\
\md_{2} \\
\vdots \\
\md_{n}
\end{array}\right),
\end{align}
which we call a data vector. Suppose our training set consists of $m_{\rm tr}$ samples scattered across parameter space, each having its own data vector. Hereafter, we will index individual data vectors across the training samples $\{1, 2, \dotsc, m_{\rm tr}\}$ with upper index $j$ such that the data vector from the $j^{\textrm{th}}$ training sample is identified as $\bd^{j}$, located in parameter space at point $\btheta^{j}$. We will also index individual data elements across the data outputs $\{1, 2, \dotsc, n\}$ with lower index $i$, such that the $i^{\textrm{th}}$ data output is identified as $\md_{i}$. The $i^{\textrm{th}}$ data output from the $j^{\textrm{th}}$ training sample is therefore uniquely identified as $\md_{i}^{j}$.

Under the standard emulator algorithm, each data output, $\md_{i}$, requires its own emulating function or predictive model. If we are only interested in a handful of outputs, then constructing an emulating function for each data output (i.e., direct emulation) is typically not hard. However, we may wish to emulate upwards of hundreds of data outputs, say for example the \tocm power spectrum at dozens of $k$ modes over dozens of individual redshifts, in which case this process becomes extremely complex. One way we can reduce this complexity is to compress our data. Instead of performing an element-by-element emulation of the data vectors, we may take advantage of the fact that different components of a data vector will tend to be correlated. For example, with the smoothness of most power spectra, neighboring $k$ and $z$ bins will be highly correlated (example \tocm power spectra are shown in \autoref{fig:21cmFAST}). There are thus fewer independent degrees of freedom than there are components in a data vector. This is the idea behind data compression techniques such as Principal Component Analysis (PCA), which seek to construct a set of principal components (PCs) that, with an appropriate choice of weights, can linearly sum to equal our data \citep{Habib2007, Higdon2008}. Transforming to the new basis of these independent modes thus constitutes a form of information compression, reducing the number of data points that must be emulated. Hereafter we will use the term principal component and eigenmode interchangeably.

To construct the principal components, we begin by taking the covariance of our training data, since it captures the typical ways in which the data vary over the parameter space. We also center the data (i.e., subtract the mean) and rescale the data (i.e., divide by a constant) such that the covariance is given by
\begin{align}
\label{eqn:PC_cov}
\bC &\equiv \left\langle\bR^{-1}\left(\bd-\overline{\bd}\right)\left(\bd-\overline{\bd}\right)^{T}\bR^{-1}\right\rangle
\end{align}
where $\overline{\bd}$ is a vector containing the average of each data output across the training set, $\bR$ is a diagonal $n\times n$ matrix containing our scaling constants, and the outer angle brackets $\langle\dotsc\rangle$ represent an average over all $m_{\textrm{tr}}$ samples in the training set. The principal components are then found by performing an eigen decomposition of the covariance matrix, given as
\begin{align}
\label{eqn:StandardPCA}
\bC\bPhi = \bPhi\bLambda,
\end{align}
where $\bPhi$ is an $n\times n$ matrix with each column representing one of the $n$ orthogonal eigenmodes (or principal components), and $\bLambda$ is a diagonal matrix containing their corresponding eigenvalues. We can think of the eigenmode matrix $\bPhi$ as a linear transformation from the basis of our centered and scaled data to a more optimal basis, given as
\begin{align}
\label{eqn:weights}
\bw^{j} &= \bPhi^{T}\left[\bR^{-1}(\bd^{j}-\overline{\bd})\right],
\end{align}
where $\bw$ is our data expressed in the new basis. This basis partitions our data into mutually exclusive, uncorrelated modes. Indeed, the covariance of our data in this basis is 
\begin{align}
\label{eqn:wwL}
\langle\bw\bw^{T}\rangle = \bLambda,
\end{align}
i.e., our eigenvalue matrix from before, which is diagonal.
We can rearrange \autoref{eqn:weights} into an expression for our original data vector, given as
\begin{align}
\bd^{j} = \overline{\bd} + \bR\bPhi\bw^{j},
\end{align}
where because $\bPhi$ is real and unitary, its inverse is equal to its transpose. This gives us insight as to why the $\bw$ vectors---the data expressed in the new basis---are called the eigenmode weights: to reconstruct our original data, we need to multiply our eigenmode matrix by an appropriate set of weights, $\bw$, and then undo our initial scaling and centering. We note that our formulation of the eigenvectors through an eigen-decomposition of a covariance matrix is similar to the approach found in \citet{Habib2007, Higdon2008, Heitmann2009}, who apply singular value decomposition (SVD) directly on the data matrix. In the case when our covariance matrix is centered and whitened (i.e., scaled by the standard deviation of the data), our two methods yield the same eigenvectors.

Although we have expressed our data in a new basis, we have not yet compressed the data because the length of $\bw^{j}$, like $\bd^{j}$, is $n$, meaning we are still using $n$ numbers to describe our data. However, one benefit of working in our new basis is that we need not use all $n$ eigenmodes to reconstruct our data vector. If we column-sort the $n$ eigenmodes in $\bPhi$ by their eigenvalues, keep those with the top $M$ eigenvalues and truncate the rest, we can approximately recover our original data vector as
\begin{align}
\label{eqn:recon}
\bd^{j} &\approx \overline{\bd} + \bR\bPhi\bw^{j},
\end{align}
where $\bPhi$ is now defined as the $n\times M$ truncated eigenmode matrix, and $\bw^j$ is now defined as the length-$M$ column vector where we have similarly sorted and then truncated the weights corresponding to the truncated eigenmodes. \emph{Hereafter, we will use $\bPhi$ and $\bw$ to exclusively mean the eigenmode matrix and weight vector respectively after truncation.} Because we are now expressing our data with $M$ numbers where $M<n$, we have compressed our data by a factor of $n/M$. The precision of this approximation depends on the inherent complexity of the training set and the number of eigenmodes we choose to keep. For our use-case, we typically achieve percent-level precision with an order-of-magnitude of compression ($n/M\sim10$).

\begin{figure}
\label{fig:eigenmodes}
\centering
\includegraphics[scale=0.7]{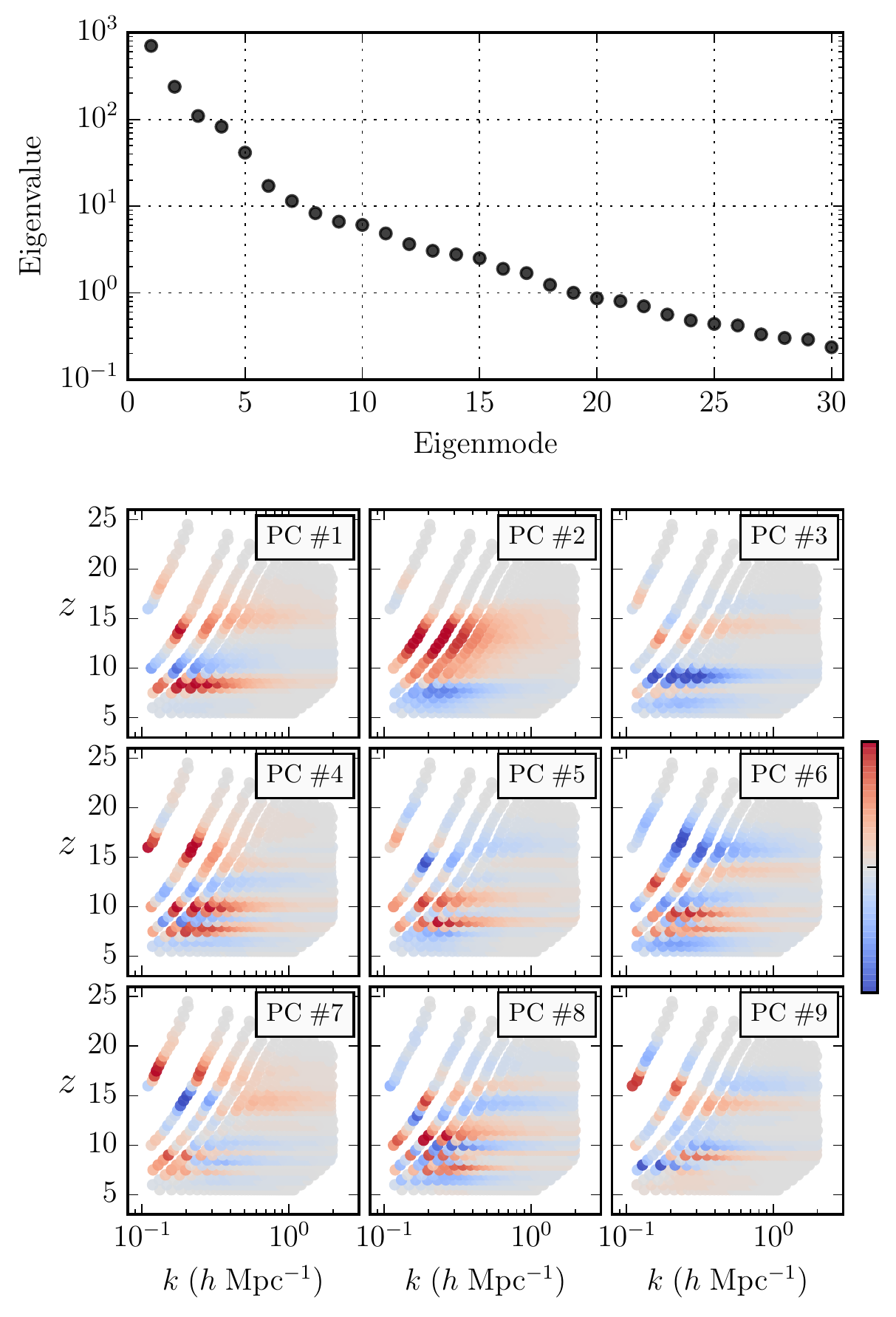}
\caption{{\bfseries Top}: Scree plot showing the eigenvalues of thirty principal components formed from training data of $\ln\DD$. {\bfseries Bottom}: The first nine principal components of the power spectrum data at each unique $k$-$z$ combination. The color scale is artificially normalized to [-1, 1] for easier comparison.}
\end{figure}

In the case where our scaling matrix, $\bR$, is the identity matrix, the formalism described above is the standard Principal Component Analysis (PCA) or Karhunen-Lo\`eve Transform (KLT). This means that PCA and KLT operate directly on the data covariance matrix formed from our unscaled data. However, not all of the $k$ modes of our power spectrum data will be measured to the same fidelity by our experiment. For the $k$ modes where our experiment will deliver higher precision measurements, our data compression technique should also yield higher precision data reconstructions. To do this, we can incorporate a non-identity scaling matrix, $\bR$, which can take an arbitrary form such that we produce eigenmodes that are desirable for the given task at hand. A natural choice would be to use the noise (or, in general, the experimental errors) of our instrument. This has the effect of downweighting portions of the data where our measurements will have minimal influence due to larger experimental errors, and conversely upweights the parts of the data with the smallest experimental errors. In the context of our worked example, we also include a whitening term in our scaling matrix, $\sigma_{\md}$, which is the standard deviation of the unscaled and centered data. After experimenting with various scaling matrices, we find a scaling matrix of $\mR_{ij} = \delta_{ij}\sigma_{\md}^{i}[\sigma_{i}/\exp(\overline{\md}_{i})]^{1/2}$ to work well, where $\delta_{ij}$ is the Kronecker delta, $\sigma$ are the observational errors, and $\exp(\overline{\md})$ is the average of the training set data, expressed in linear (not logarithmic) space.

An example set of principal components formed from our training data is shown in \autoref{fig:eigenmodes}, where we display the first nine principal components (eigenmodes) of the log, centered and scaled $\DD$ training data. We discuss the simulation used to generate this training data in \autoref{sec:simulations}. The amplitude of the PCs have been artificially normalized to unity for easier comparison. We find in general that at a particular redshift, an individual PC tends to be smooth and positively correlated along $k$, and at a particular $k$ shows negative and positive correlations across redshift. This is a reflection of the underlying smoothness of the power spectra across $k$, and the fact that physical processes such as reionization, X-ray heating and Lyman-$\alpha$ coupling tend to produce redshift-dependent peaks and troughs in the power spectrum (see e.g., \autoref{fig:21cmFAST}). The reason why the PCs lose strength at high $k$ is because our rescaling matrix $\bR$ downweights our data covariance matrix at high $k$. As we will see in \autoref{sec:sensitivity}, the bulk of a \tocm experiment's sensitivity to the power spectrum is located at lower $k$.

\subsection{Gaussian Process Regression}
\label{sec:GPR}

For the purposes of emulation, we are not interested in merely reconstructing our original training set data at their corresponding points in parameter space $\btheta^j$, but are interested in constructing a prediction of a \emph{new} data vector, $\bd^{\rm new}$, at a new position in our parameter space, $\btheta^{\rm new}$. We can construct a prediction of the data vector at this new point in parameter space by evaluating \autoref{eqn:recon} with $\bw^{\rm new}$; however, we do not know this weight vector a priori. To estimate it at any point in parameter space, we require a predictive function for each element of $\bw$ spanning the entire parameter space. In other words, we need to interpolate $\bw$ over our parameter space. To do this, we adopt a Gaussian Process (GP) model, which is a highly flexible and non-parametric regressor. See \citet{Rasmussen2006} for a review on Gaussian Processes for regression.

A GP is fully specified by its mean function and covariance kernel. The GP mean function can be thought of as the global trend of the data, while its covariance kernel describes the correlated random Gaussian fluctuations about the mean trend. In practice, because we center our data about zero before constructing the principal components and their weights (\autoref{eqn:PC_cov}), we set our mean function to be identically zero. For the covariance kernel we employ a standard squared-exponential kernel, which is fully stationary, infinitely differentiable, produces smooth realizations of a correlated random Gaussian field, and is given in multivariate form as
\begin{align}
\label{eqn:kernel}
k(\btheta,\btheta^{\prime}|\bL) &= \sigma_{\mA}^{2}\cdot\exp\left[-\frac{1}{2}\left(\btheta-\btheta^{\prime}\right)^{T}\bL^{-2}\left(\btheta-\btheta^{\prime}\right)\right],
\end{align} 
where $\btheta$ and $\btheta^{\prime}$ denote two position vectors in our parameter space, $\bL$ is a diagonal matrix containing the characteristic scale length of correlations $\ell$ across each parameter, and $\sigma_{\mA}$ is the characteristic amplitude of the covariance. $\bL$ is a tunable hyperparameter of the kernel function that must be selected \emph{a priori}. We discuss how we make these choices in \autoref{sec:hyper_params}. We set $\sigma_{\mA} = 1$ and therefore it is not a hyperparameter of our kernel. For this to be valid, we must scale the eigenmode weight training data to have variance of unity.

In our case, we have multiple GP regressors---one for each component of the eigenmode weight vector. Consider for example the weight for the first eigenmode. Suppose we group the training data for this weight into a vector $\by^{\text{tr}}$, such that $y^{\text{tr}}_j \equiv w_1^j / \lambda_{1}^{1/2}$, where $\lambda_{i}$ is the variance of weight element $\mw_{i}$ from \autoref{eqn:wwL}. Dividing by the standard deviation ensures that the variance of the weights are unity, and therefore allows us to set $\sigma_{\mA} = 1$. If we define an $m_\textrm{tr} \times  m_\textrm{tr}$ matrix $\bK^{\textrm{tr-tr}}_{1}$ such that $(\mK^{\textrm{tr-tr}}_{1})_{ij}~\equiv~k(\btheta^\textrm{tr}_i,\btheta^\textrm{tr}_j|\bL_{1})$, then the GP prediction for the weight at point $\btheta^{\rm new}$ is given by
\begin{align}
\label{eqn:pred_fun}
w^{\rm new}_1 = \lambda_{1}^{1/2}(\bk^{{\rm new}\textrm{-tr}}_{1})^T \left[\bK^{\textrm{tr-tr}}_{1}+\sigma_{n}^{2}\bI\right]^{-1}\by^{\text{tr}},
\end{align}
where $\bk^{{\rm new}\textrm{-tr}}_{1}$ is a length-$m_\textrm{tr}$ vector defined analogously to $\bK^{\textrm{tr-tr}}_{1}$, i.e., $(\mk^{{\rm new}\textrm{-tr}}_{1})_{i}~\equiv~k(\btheta^{\rm new},\btheta^\textrm{tr}_{i}|\bL_{1})$, $\bL_{1}$ is the matrix containing the hyperparameters chosen a priori for the input training data and the subscript 1 specifies that the input training data are the weights of the first PC mode, $w_{1}$. The variance about this prediction is then given by\footnote{In principle, one may perform a GP estimate over several points in parameter space at once. \autoref{eqn:pred_fun} then predicts an entire vector of $w_1^{\rm new}$ values simultaneously, and \autoref{eqn:pred_cov} generalizes to a full covariance matrix. Here we do not employ such a formalism since an MCMC chain explores parameter space one point at a time.}
\begin{align}
\label{eqn:pred_cov}
\begin{split}
\gamma^{{\rm new}}_1 =\ & 1 - (\bk^{{\rm new}\textrm{-tr}}_{1})^T\left(\bK^{\textrm{tr-tr}}_{1} + \sigma_{n}^{2}\bI\right)^{-1}\bk^{{\rm new}\textrm{-tr}}_{1},
\end{split}
\end{align}
where $\bI$ is the identity matrix, and $\sigma_{n}^{2}$ is the variance of random Gaussian noise possibly corrupting the training data from their underlying distribution and is a hyperparameter of the GP \citep{Rasmussen2006}.

Evaluating \autoref{eqn:pred_fun} for each PC weight yields a set of predicted weights that come together to form the vector $\mathbf{w}^{\rm new}$. This may then be inserted into \autoref{eqn:recon} to yield predictions for the quantities we desire. Similarly, evaluating \autoref{eqn:pred_cov} for each PC weight and stacking them into a vector $\bgamma^{\rm new}$, we may propagate our GP's uncertainty on $\bw^{\rm new}$ into an emulator covariance $\bSigma_{\mE}$, which describes the uncertainty on the unlogged\footnote{Recall that in \autoref{eq:ddef} we defined the data vector to be the \emph{logarithm} of the original quantities we wished to emulate.} emulator predictions $\exp(\bd^{\rm new})$, and is given by
\begin{align}
\label{eqn:data_recon_cov}
(\Sigma_{\mE})_{ij} &=  \sum_{k}^{M} \exp(\md^{\rm new}_i) \exp(\md^{\rm new}_j) \Phi_{ik} \Phi_{jk}\gamma_{k}^{{\rm new}},
\end{align}
where in deriving this expression we have assumed that the emulator errors are small. Importantly, note that because $\gamma_{k}^{{\rm new}}$ depends on $\btheta^{\rm new}$, the same is true for $\bSigma_{\mE}$. This is to be expected. For instance, one would intuitively expect the emulator error to be larger towards the edge of our training region than at the center of it. In practice, it is helpful to complement estimates of emulator from \autoref{eqn:data_recon_cov} with empirical estimates derived from cross validation. Essentially, one takes a set of simulation evaluations not in the training set and compares the emulator's prediction at those points in parameter space against the true simulation output. We further discuss these considerations and how the estimated emulator error $\bSigma_{\mE}$ comes into our parameter constraints when we lay out the construction of our likelihood function in \autoref{sec:likelihood}.

So far we have been working towards constructing a set of GP models for each PC mode, each of which is a predictive function spanning the entire parameter space and uses all of the training data. A different regression strategy is called the Learn-As-You-Go method \citep{Aslanyan2015}. In this method, one takes a small subset of the training data immediately surrounding the point-of-interest, $\btheta^{\rm new}$, in order to construct localized predictive functions, which then get thrown away after the prediction is made. This is desirable when the training set becomes exceedingly large ($m_{\rm tr} \gtrsim 10^{4}$ samples), because the computational cost of GP regression naively scales as $m_{\rm tr}^{3}$. This is the regression strategy we adopt in our initial broad parameter space search in \autoref{sec:param_explore}.

Our emulator algorithm in \texttt{emupy} relies on base code from the Gaussian Process module in the publicly-available Python package Sci-Kit Learn,\footnote{\url{http://scikit-learn.org/}} which has an optimized implementation of \autoref{eqn:pred_fun} and \autoref{eqn:pred_cov} \citep{Pedregosa2012}.

\subsubsection{GP Hyperparameter Solution}
\label{sec:hyper_params}

The problem we have yet to address is how to select the proper set of hyperparameters for our GP kernel function, in particular the characteristic scale length of correlations $\ell$ across each model parameter. We can do this through a model selection analysis, where we seek to find $\bL$ such that the marginal likelihood of the training data given the model hyperparameters is maximized. From \citet{Rasmussen2006}, the GP log-marginal likelihood for a single PC mode is given (up to a constant) by
\begin{align}
\ln\mcL_{\text{M}}(\by^{\text{tr}}|\bL) &\propto -\frac{1}{2}(\by^{\text{tr}})^{T}(\bK^{\textrm{tr-tr}})^{-1}\by^{\text{tr}}-\frac{1}{2} \textrm{det}(\bK^{\text{tr-tr}}),
\end{align}
where $\bK^{\rm tr\textrm{-tr}}$ has the same definition as in \autoref{eqn:pred_fun}, and thus carries with it a dependence on $\btheta^{\rm tr}$ and $\bL$. In principle, one could also simultaneously vary the assumed noise variance ($\sigma_{n}^{2}$) of the target data as an additional hyperparameter and jointly fit for the combination of $[\bL, \sigma_{n}^{2}]$. To find these optimal hyperparameters, we can use a gradient descent algorithm to explore the hyperparameter parameter space of $\bL$ and $\sigma_{n}^{2}$ until we find a combination that maximizes $\ln\mcL_{\text{ML}}$. When working with training data directly from simulations, we would expect $\sigma_{n}^{2}$ to be minimal; we are not dealing with any observational or instrumental systematics that might introduce uncertainty into their underlying values. Depending on the simulation, there may be numerical noise or artifacts that introduce excess noise or outlier points into the target data, which may skew the resultant best-fit for $\bL$ or break the hyperparameter regression entirely. This can be alleviated by keeping $\sigma_{n}^{2}$ as a free parameter and fitting for it and $\bL$ jointly.

In our eleven dimensional space, this calculation can become exceedingly slow when the number of samples in our training set exceeds ten thousand. In our initial broad parameter space exploration (\autoref{sec:param_explore}), for example, performing a hyperparameter gradient descent with all 15,000 samples is not attempted. To solve for the hyperparameters, we thus build a separate training set that slices the parameter space along each parameter and lays down samples along that parameter while holding all others constant. We then take this training set slice and train a 1D GP and fit for the optimal $\ell$ of that parameter by maximizing the marginal likelihood. We repeat this for each parameter and then form our $\bL$ matrix by inserting our calculated $\ell$ along its diagonal. This is a method of constraining each dimension's $\ell$ individually, in contrast to the previous method of constraining $\ell$ across all dimensions jointly. While this is a more approximate method, it is computationally much faster.

In order to construct a fully hierarchical model, we should in principle not be selecting a single set of hyperparameters for our GP fit, but instead should be marginalizing over all allowed hyperparameter combinations informed by the marginal likelihood. That is to say, we should fold the uncertainty on the optimal choice of $\ell$ into our uncertainty on our predicted $\bw^{{\rm new}}$ and thus our predicted $\bd^{{\rm new}}$. In theory this would be ideal, but in practice, this quickly becomes computationally infeasible. This is because the time it takes to train a GP and make interpolation predictions naively scales as the number of training samples $m_{\rm tr}$ cubed. Optimized implementations, such as the one in Sci-Kit Learn, achieve better scaling for low $m_{\rm tr}$, but for large $m_{\rm tr}$ this efficiency quickly drops, to the point where having to marginalize over the hyperparameters to make a single prediction at a single point in parameter space can take upwards of minutes, if not hours, which begins to approach the run time of our original simulation. Furthermore, all of these concerns are exponentially exacerbated in high dimensional spaces. However, in the limit of a high training set sampling density this difference becomes small, which is to say that our marginal likelihood becomes narrow as a function of the hyperparameters. Lastly, and most importantly, we can always turn to diagnose the accuracy of our emulator (and calibrate out its failures) by cross validating it against a separate set of simulation evaluations. In doing so, we can ensure that the emulator is accurate within the space enclosed by our training set.

\subsubsection{GP Cross Validation}
\label{sec:cross_valid}

Emulators are approximations to the data products of interest, and as such we need to be able to assess their performance if we are to trust the parameter constraints we produce with them. As discussed above, this can be accomplished empirically via cross validation (CV). In this paper, rather than computing extra simulation outputs to serve as CV samples, we elect to perform $k$-fold cross validation. In $k$-fold cross validation, we take a subset of our training samples and separate them out, train our emulator on the remaining samples, cross validate over the separated set, and then repeat this $k$ times. This ensures we are not training on the cross validation samples and also means we do not have to use extra computational resources generating new samples.

\begin{figure}
\label{fig:cv_plots}
\centering
\includegraphics[scale=0.7]{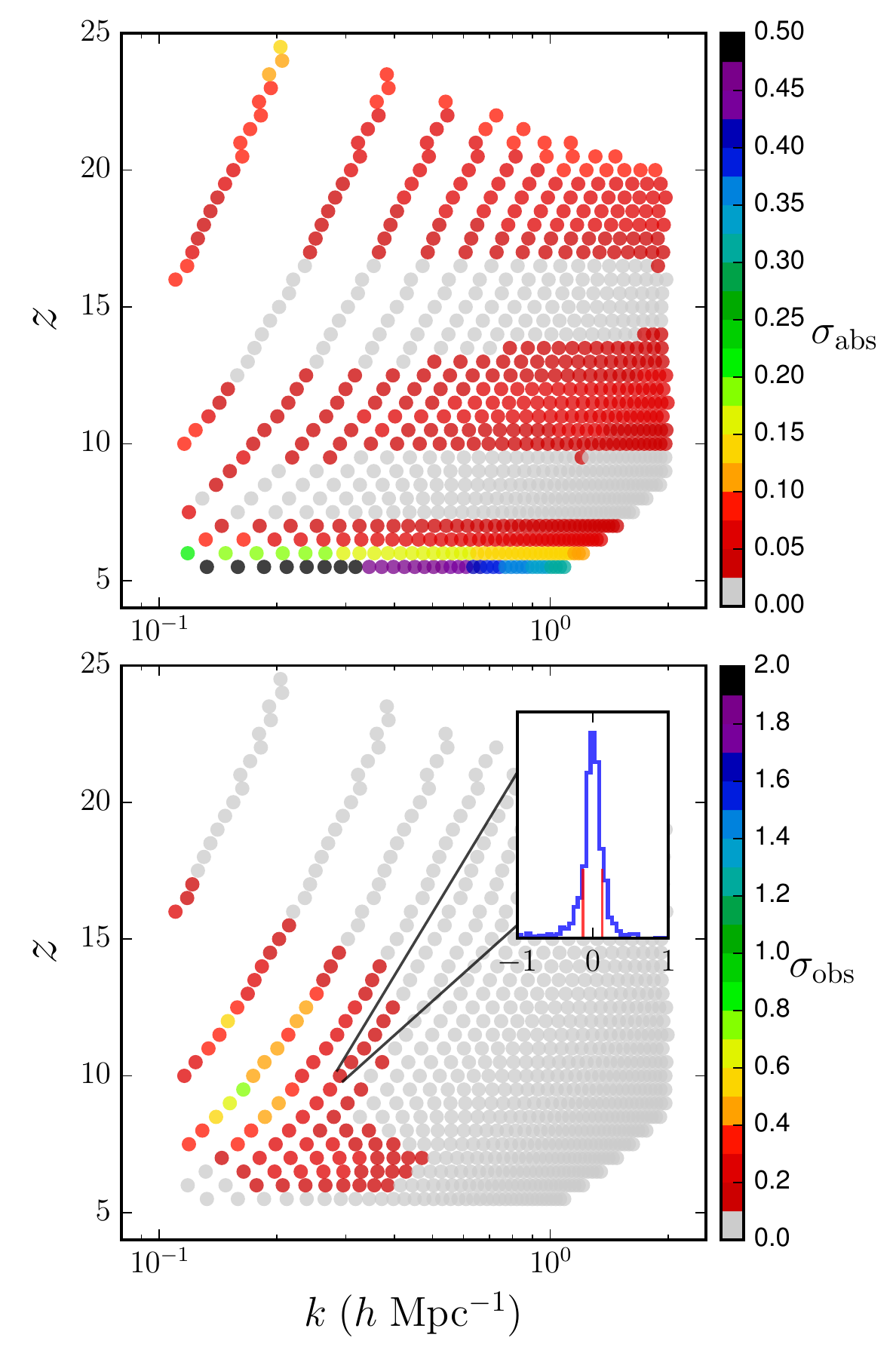}
\caption{{\bfseries Top}: Standard deviation of the absolute fractional emulator error ($\sigma_{\mathrm{abs}}$) with respect to the CV set. Grey color indicates an emulator precision of $\le2.5\%$. {\bfseries Bottom}: Standard deviation of the offset between emulator prediction and CV data, divided by the experimental errors ($\sigma_{\mathrm{obs}}$). The grey color over the majority of the data signifies we can recover the data to $\le10\%$ relative to the experimental error bars. {\bfseries Inset}: Error distribution $\epsilon_{\mathrm{obs}}$ for a data output, with its robust standard deviation marked as vertical bars.}
\end{figure}

We use two error metrics to quantify the emulator performance. The first metric is the absolute fractional error of the emulator prediction over the cross validation data, expressed as $\epsilon_{\mathrm{abs}} = ([\DD]_{\mE}-[\DD]_{\mathrm{CV}})/[\DD]_{\mathrm{CV}}$. This gives us a sense of the absolute precision of our emulator. However, not all $k$ modes contribute equally to our constraining power. Because our \tocm experiment will measure some $k$ modes to significantly higher signal-to-noise (S/N) than other $k$ modes, we want to confirm that our emulator can at least emulate at a precision better than the S/N of our experiment, and is therefore not the dominant source of error at those $k$ modes. The second metric we use is then the offset between the emulator and the CV, divided by the error bars of our \tocm experiment, given by $\epsilon_{\mathrm{obs}} \equiv ([\DD]_{\mE}-[\DD]_{\mathrm{CV}})/\sigma_{\mS}$, where $\sigma_{\mS}$ are the \tocm power spectrum error bars of our experiment. For this, we use the projected sensitivity error bars of the HERA experiment, which we discuss in detail in \autoref{sec:sensitivity}, and is shown in \autoref{fig:mock_obs}.

Cross validation leaves us with an error distribution of the CV samples at each unique $k$ and $z$. Applying our error metrics, we are left with two sets of error distributions for the emulated power spectra, $\epsilon_{\mathrm{abs}}(k,z)$ and $\epsilon_{\mathrm{obs}}(k,z)$. To quantify their characteristic widths, we calculate their robust standard deviations, $\sigma_{\mathrm{abs}}$ and $\sigma_{\mathrm{obs}}$ respectively, using the bi-weight method of \citet{Beers1990}. We show an example of these standard deviations in \autoref{fig:cv_plots}, which demonstrates the emulator's ability to recover the \tocm power spectra having trained it on the training set described in \autoref{sec:constraints}. Here, we take $2\times10^{3}$ of the center-most samples of the $5\times10^{3}$-sample training set and perform 5-fold cross validation on them. The top subplot shows the absolute error metric $\sigma_{\mathrm{abs}}$, and demonstrates our ability to emulate at $\le5\%$ precision for the majority of the power spectra, and $\le10\%$ for almost all of the power spectra. The bottom subplot shows the observational error metric $\sigma_{\mathrm{obs}}$, and demonstrates that we can emulate to an average precision that is well below the observational error bars of a HERA-like experiment for virtually all $k$ modes, keeping in mind that the highest S/N $k$ modes for \tocm experiments are at low-$k$ and low-$z$ for $z\gtrsim6$. The inset shows the underlying error distribution and its robust standard deviation for one of the power spectra data output. 
Note that the the distribution of points on the $k$-$z$ plane that are shown in \autoref{fig:cv_plots} are not determined by the emulator; indeed, one can easily emulate the power spectra at different values of $k$ and $z$. Instead, these points were chosen to match the observational survey parameters and our choice of binning along our observational bandpass. We discuss such survey parameters in more detail in \autoref{sec:sensitivity}.

The observational error metric is of course dependent on the chosen \tocm experiment and its power spectrum sensitivity. This particular emulator design, for example, may not be precise enough to emulate within the error bars of a futuristic experiment. If we need to boost our emulator's precision, we can do so to an almost arbitrary level by simply generating more training samples and packing the space more densely. The limiting factors of this is the need to generate an additional number of training samples which is unknown a priori, and the intrinsic $m_{\text{tr}}^{3}$ scaling of the Gaussian Process regressor. However, with sufficient computational resources and novel emulation strategies like Learn-As-You-Go, increasing the emulator's precision to match an arbitrary experimental precision is in principle feasible.


\section{Choosing a Model for Cosmic Dawn}
\label{sec:simulations}
Having described the core features of our emulator, we will now focus on a specific model of the \tocm signal so that we may build a training set. To accurately describe the large-scale correlation statistics of the cosmological \tocm signal, we need large-volume simulations with box lengths $L > 200\,\textrm{Mpc}$ \citep{Barkana2004,Trac2011,Iliev2014}. Compared to the physical sizes of ionizing photon sources and sinks at the galactic scale of kpc and smaller, it is clear that in order to directly simulate reionization one needs to resolve size scales that span many orders of magnitude. This has made direct simulations of reionization a computationally formidable task; current state-of-the-art high resolution hydrodynamic plus self-consistent radiative transfer codes are extremely expensive and only reach up to tens of Mpc in box length. As a consequence, less numerically rigorous but dramatically cheaper semi-analytic approaches have made more progress in exploring the EoR parameter space. One such code is the semi-numerical simulation \texttt{21cmFAST} \citep{Mesinger2007, Mesinger2011}, which we use in this work to build our training sets. For the following, we use the publicly available \texttt{21cmFAST\_v1.12}.\footnote{\url{https://github.com/andreimesinger/21cmFAST}} However, we again emphasize that the idea of emulating simulations of Cosmic Dawn is not one that is tied to \texttt{21cmFAST}; indeed, one could easily perform similar calculations to the one in this paper with other semi-numerical simulations, such as those described in \citet{Geil2008}, \citet{Choudhury2009}, \citet{Thomas2009}, \citet{Santos2010}, \citet{Battaglia2013}, and \citet{Kulkarni2016}, or with numerical simulations, such as those described in \citet{Mellema2006}, \citet{Zahn2007}, \citet{Baek2009}, \citet{Trac2011}, \citet{Iliev2014}, \citet{Gnedin2014}, \citet{Ross2016}, \citet{Kaurov2016}, and \citet{Das2017}.

\begin{figure*}
\label{fig:21cmFAST}
\centering
\includegraphics[scale=0.65]{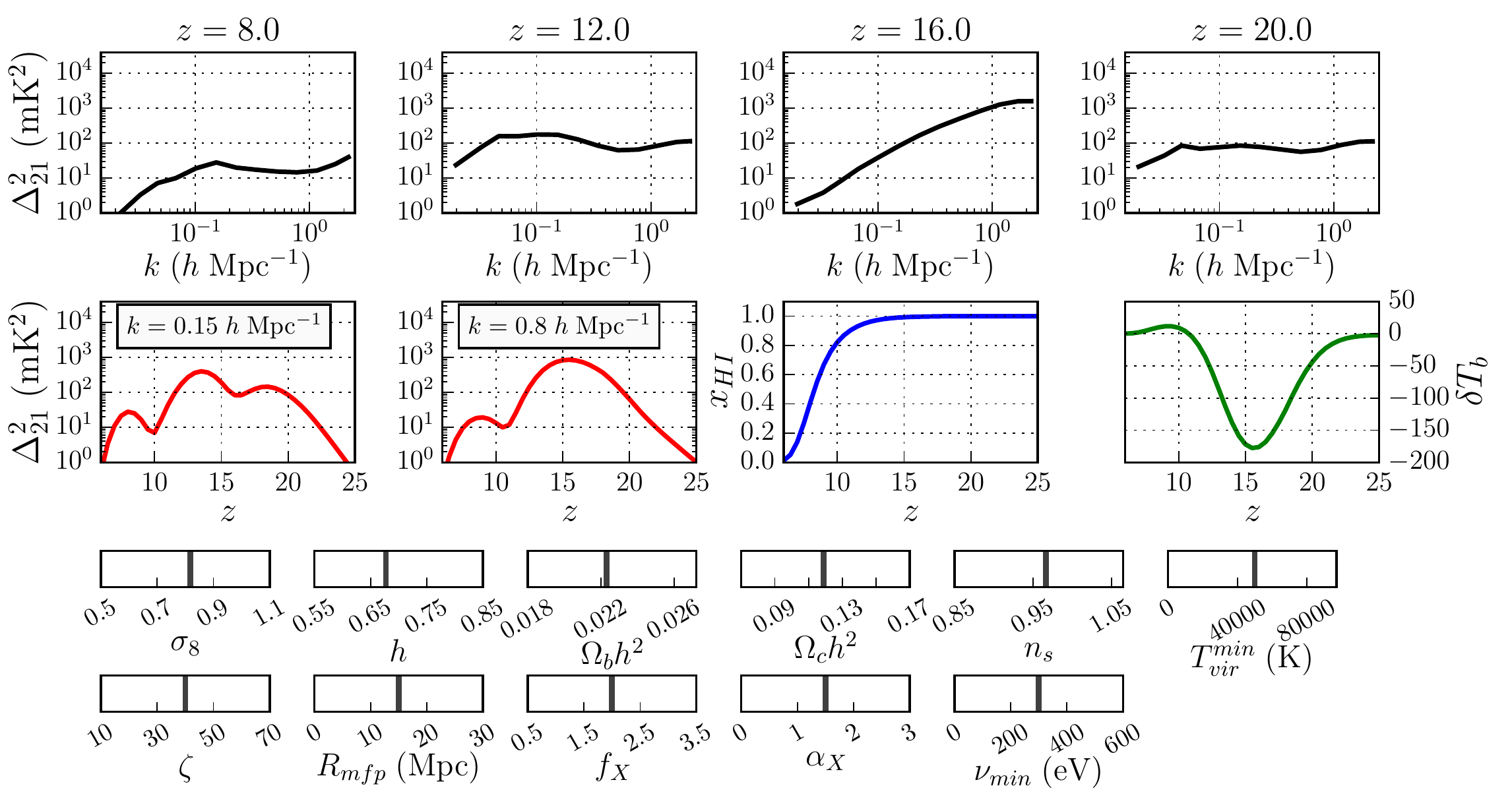}
\caption{Data products from the semi-numerical EoR simulation \texttt{21cmFAST}. From top to bottom, left to right we show the \tocm dimensional power spectra as a function of wavenumber $k$ at various redshifts, the power spectra redshift evolution at a specific $k$-mode, the hydrogen neutral fraction redshift evolution and the global signal redshift evolution. The parameters on bottom show the choice of model parameters for this specific realization. For building intuition, a movie showing the behavior of these outputs to variations in the model parameters can be found at \url{http://w.astro.berkeley.edu/~nkern/images/ps_movie.mp4}.}
\end{figure*}

The relevant observable our simulation needs to predict is the \tocm brightness temperature at radio frequencies. Specifically, it is the \tocm brightness temperature offset from the background CMB brightness temperature. Because the \tocm signal is a line transition, its fluctuations across frequency encode redshift information while fluctuations across the sky encode angular information. We can therefore recover three-dimensional information of the IGM structure and thermal state with the \tocm line. For a parcel of gas at a specific location on the sky with a redshift $z$ corresponding to a redshifted \tocm frequency of $\nu$, the \tocm brightness temperature offset can be written as
\begin{align}
\label{eqn:dTb}
\Tb(\nu) \approx 9\xHI(1+\delta)(1+z)^{\frac{1}{2}}\left(1-\frac{\Tg}{\Ts}\right)\left(\frac{H(z)}{d v_{r}/dr}\right) \text{mK}
\end{align}
where $\xHI$ is the hydrogen neutral fraction, $\delta$ the baryon overdensity, $\Tg$ the CMB background temperature, $\Ts$ the neutral hydrogen hyperfine ``spin'' temperature \citep{Wouthuysen1952, Field1958, Furlanetto2006a}, $H(z)$ is the Hubble parameter, and $dv_{r}/dr$ is the line of sight proper velocity gradient. All of the parameters on the right hand side have both frequency and angular dependence on the sky, except for the Hubble parameter with only frequency dependence. To make a prediction of the \tocm brightness temperature field, we therefore need an underlying cosmology, a prescription for the matter density field, the hydrogen ionization fraction field and in certain cases the hyperfine spin temperature field. Some models choose to make the assumption that the spin temperature greatly exceeds the photon temperature ($\Ts \gg \Tg$), in which case the \tocm temperature is insensitive to $\Ts$ and we need not compute it. This is sometimes assumed to be true during the late stages of reionization ($z<10$) when the IGM gas temperature has been sufficiently heated. This is complicated by the fact that certain EoR scenarios of reionization, dubbed``cold reionization,'' predict inefficient IGM heating and therefore the $\Ts\gg\Tg$ assumption breaks down \citep{Mesinger2014, Cohen2016, Mirocha2017}. Furthermore, recent work shows that even if the spin temperature is saturated with respect to the photon temperature at $z<10$, efficient IGM heating can still leave an imprint on EoR and will bias astrophysical constraints that neglect it \citep{Greig2017b}. 

\texttt{21cmFAST} generates a density field by evolving an initial Gaussian random field to low redshifts via the Zel'dovich approximation. \texttt{21cmFAST} does not account for baryonic hydrodynamics and thus makes the implicit assumption that the baryons track the dark matter. From the evolved density field, hydrogen ionization fields are calculated using the excursion set theory of \citet{Furlanetto2004}. In this formalism, the density field is smoothed to a comoving radius scale, $\Rmfp$, and the central cell about the smoothing is considered ionized if
\begin{align}
\fcoll(\bm{x}, z, R) \ge \zeta^{-1},
\end{align}
where $\fcoll$ is the fraction of matter that has collapsed into gravitationally bound structure at position $\bm{x}$, redshift $z$ and with smoothing scale $R$. $\fcoll$ is computed via the hybrid prescription in \citet{Barkana2004}. $\zeta$ is an ionization efficiency parameter for star forming halos (see \autoref{sec:eor_params} below for a description). This process is iterated on smaller smoothing scales until either the cell becomes ionized or the cell resolution is reached. The initial smoothing scale, $\Rmfp$, can be thought of as the mean-free path of photons through ionized regions, which accounts for unresolved sub-halos with pockets of neutral hydrogen that act as ionization sinks. Detailed studies have compared \texttt{21cmFAST} against more accurate RT simulations and have shown that their ionization fields and \tocm power spectra during the EoR ($z<10$) are consistent at the $\sim20\%$ level \citep{Mesinger2011, Zahn2011}.

\texttt{21cmFAST} can also calculate the kinetic gas temperature and spin temperature fields. The spin temperature couples to either the background photon CMB temperature or to the IGM kinetic gas temperature. The latter can occur via hydrogen collisional coupling, as is thought to occur early on ($z\ge20$), or via the Wouthuysen--Field effect where Lyman-$\alpha$ pumping couples the spin temperature to the Lyman-$\alpha$ color temperature, which closely traces the kinetic gas temperature due to the high scattering cross section of Lyman-$\alpha$ photons with neutral hydrogen \citep{Furlanetto2006b}. In order to calculate the IGM gas kinetic temperature one must track inhomogeneous IGM heating, which is thought to predominately occur by X-rays. To track this, \texttt{21cmFAST} integrates the angle-averaged specific X-ray emissivity across a light cone and across X-ray frequencies for each cell. X-ray production, due to either high-mass X-ray binaries or a hot Interstellar Medium (ISM) phase, is assumed to be tied to star formation. While the power spectra during X-ray heating from a fiducial \texttt{21cmFAST} realization roughly agree with the trends seen from numerical simulations \citep{Mesinger2011, Ross2016}, a detailed comparison of \texttt{21cmFAST} against numerical simulations of X-ray heating has not been made. Such comparisons are necessary to test the accuracy of the X-ray treatment in \texttt{21cmFAST}, and could help calibrate or better inform the semi-analytics therein. For the time being, we accept \texttt{21cmFAST}'s treatment of the X-ray heating for intuitive purposes. For a more detailed description of the semi-numerics inside \texttt{21cmFAST}, see \citet{Mesinger2011, Mesinger2013}.

To build our training sets, the simulation runs have box lengths $L=400$ Mpc. We sample the Gaussian initial conditions for the density field from a $800^{3}$-voxel grid, which then get smoothed onto a $200^{3}$-voxel grid to track its evolution via the Zel'dovich approximation and to compute the relevant \tocm fields. This lower resolution grid corresponds to a cell resolution of 2 Mpc. For comparison, the minimum length-scale that an experiment like HERA is expected to be sensitive to is around 5 Mpc.

The data products that \texttt{21cmFAST} produces are 3D box outputs of cosmological fields, from which we can construct the \tocm power spectrum, the average \tocm brightness temperature offset from the CMB (also referred to as the global signal or monopole signal), the average hydrogen neutral fraction and the integrated electron scattering optical depth. \autoref{fig:21cmFAST} shows an example of these data products from a fiducial \texttt{21cmFAST} realization. We could also construct higher order statistical probes from the box-outputs, such as three- or four-point functions, which in principle carry useful information because the ionization fields are non-Gaussian; however, for this study we focus on the power spectrum as our observable. Future work will focus on synthesizing other EoR probes within the parameter estimation framework presented here.

The \tocm power spectrum is defined as $\DD(k) = (k^{3}/2\pi^{2})P_{21}(k)$, with $P_{21}(k)$ being defined as
\begin{align}
\label{eqn:DD}
\langle\widetilde{\Tb}(\bm{k}_{1})\widetilde{\Tb}^{\ast}(\bm{k}_{2})\rangle = (2\pi)^{3}\delta^{D}(\bm{k}_{1}-\bm{k}_{2})P_{21}({k}_{1}),
\end{align}
where $\langle\dotsc\rangle$ denotes an ensemble average, $\widetilde{\Tb}(\bm{k})$ is the spatial Fourier transform of $\Tb(\bm{x})$, $\delta^{D}$ is the Dirac delta function, $\bm{k}$ is the spatial Fourier wavevector, and $k\equiv|\bm{k}|$. Because we constructed the power spectrum by taking the spatial Fourier transform of $\Tb$, $\DD$ carries units of mK$^{2}$. This is the statistic \tocm interferometric experiments like HERA are aiming to measure (among other quantities), and this is the \tocm statistic we will focus on in this paper.

\subsection{Model Parameterization}
\label{sec:parameterization}

We adopt an eleven parameter model within \texttt{21cmFAST} to characterize the variability of $\Tb$ across the reionization and X-ray heating epochs. This consists of six astrophysical parameters that describe the production and propagation of UV and X-ray photons, and five cosmological parameters that influence the underlying density field and velocity fields of our Universe. 

\subsubsection{EoR Parameters: $\zeta, \Tvirmin, \Rmfp$}
\label{sec:eor_params}

The production rate of UV photons is governed by the ionization efficiency of star-forming galaxies, $\zeta$, which can be expressed as
\begin{align}
\label{eqn:zeta}
\zeta = 30\left(\frac{\fesc}{0.15}\right)\left(\frac{f_{\star}}{0.1}\right)\left(\frac{N_{\gamma}}{4000}\right)\left(\frac{2}{1+n_{\text{rec}}}\right),
\end{align}
where $\fesc$ is the fraction of produced UV photons that escape the galaxy, $f_{\star}$ is the fraction of collapsed gas in stars, $N_{\gamma}$ is the number of ionizing photons produced per stellar baryon and $n_{\text{rec}}$ is the average number of times a hydrogen atom in the IGM recombines. The splitting of $\zeta$ into these four constituent parameters is merely for clarity: the numerics of \texttt{21cmFAST} respond only to a change in $\zeta$, regardless of what sub-parameter sourced that change. These sub-parameters are therefore completely degenerate with each other in the way they affect reionization in \texttt{21cmFAST}. Previous works have explored how to parameterize the mass and redshift evolution of $\zeta$ \citep{Greig2015a, Sun2016}, and this will certainly be a feature to incorporate into this framework for future studies. For the time being, we assume $\zeta$ to be a constant for intuitive purposes, similar to previous work. Some of the fiducial values for the terms in \autoref{eqn:zeta} are physically motivated---$N_{\gamma} \sim4000$ is the expectation from spectral models of Population II stars \citep{Barkana2005}, and both $f_{\star}$ and $f_{\text{esc}}$ are thought to lie within a few factors of 0.1 \citep{Kuhlen2012, Robertson2015, Paardekooper2015, Xu2016, Sun2016}---however, these are not strongly constrained at high redshifts and are particularly unconstrained for low-mass halos.

Baryonic matter must cool in order for it to condense and allow for star formation. This can occur through radiative cooling from molecular hydrogen, although this is easily photodissociated by Lyman-Werner photons from stellar feedback \citep{Haiman1997}. Other cooling pathways exist, but in general, low mass mini-halos are thought to have poor star formation efficiencies due to stellar feedback \citep{Haiman2000}. We can parameterize the lower limit on halo mass for efficient star formation as a minimum halo virial temperature, $\Tvirmin$ (K). Here we adopt a fiducial $\Tvirmin$ of $5\times10^{4}$ K, above the atomic line cooling threshold of $10^{4}$ K \citep{Barkana2002}. In practice, this parameter has the effect of stopping the excursion set formalism for a cell smoothed on scale $R$ if its mass is less than the minimum mass set by $\Tvirmin$.

As ionizing photons escape star forming galaxies and propagate through their local HII region, they are expected to encounter pockets of neutral hydrogen in highly shielded sub-structures (Lyman-limit systems). Without explicitly resolving these ionization sinks, we can parameterize their effect on ionizing photons escaping a galaxy by setting an effective mean-free path through HII regions for UV photons, $\Rmfp$. In practice, this sets the maximum bubble size around ionization sources, and is the initial smoothing scale for the excursion set (as discussed above in \autoref{sec:simulations}). Motivated by subgrid modeling of inhomogeneous recombinations \citep{Sobacchi2014}, we adopt a fiducial value of $\Rmfp = 15\,\textrm{Mpc}$.

\subsubsection{X-ray Spectral Parameters: $\fX$, $\aX$, $\numin$}
The sensitivity of the \tocm power spectrum to cosmic X-rays during the IGM heating epoch may allow us to constrain the spectral properties of the X-ray generating sources. These are theorized to come predominately from either High Mass X-ray Binaries (HMXB) or a hot Interstellar Medium (ISM) component in galaxies heated by supernovae. In \texttt{21cmFAST}, the X-ray source emissivity is proportional to
\begin{align}
\epsilon_{X}(\nu) \propto \fX\left(\frac{\nu}{\numin}\right)^{-\aX},
\end{align}
where $\fX$ is the X-ray efficiency parameter (an overall normalization), $\aX$ is the spectral slope parameter, and $\numin$ is the obscuration frequency cutoff parameter, below which we take the X-ray emissivity to be zero due to ISM absorption. High-resolution hydrodynamic simulations of the X-ray opacity within the ISM have found that such a power-law model is a reasonable approximation of the emergent X-ray spectrum from the first galaxies \citep{Das2017}. Our fiducial choice of $\fX = 1$ corresponds to an average of 0.1 X-ray photons produced per stellar baryon. HMXB spectra have typical spectral slopes $\aX$ of roughly unity, while a hot ISM component tends to have a spectral slope of roughly 3 \citep{Mineo2012}. Our fiducial choice of $\aX=1.5$ straddles these expectations. The obscuration cutoff frequency, $\numin$, parameterizes the X-ray optical depth of the ISM in the first galaxies and is dependent on their column densities and metallicities. We choose a fiducial value of $\numin = 0.3\,\textrm{keV}$, consistent with previous theoretical work \citep{Pacucci2014, Ewall-Wice2016a}. Because the model assumes the X-ray production comes from star forming halos, the EoR parameter $\Tvirmin$ also affects the spatial distribution of X-ray sources, and is therefore also implicitly an X-ray heating parameter. For a more detailed description of the X-ray numerics in \texttt{21cmFAST}, see \citet{Mesinger2013}.

\begin{figure*}
\label{fig:mock_obs}
\centering
\includegraphics[scale=1.0]{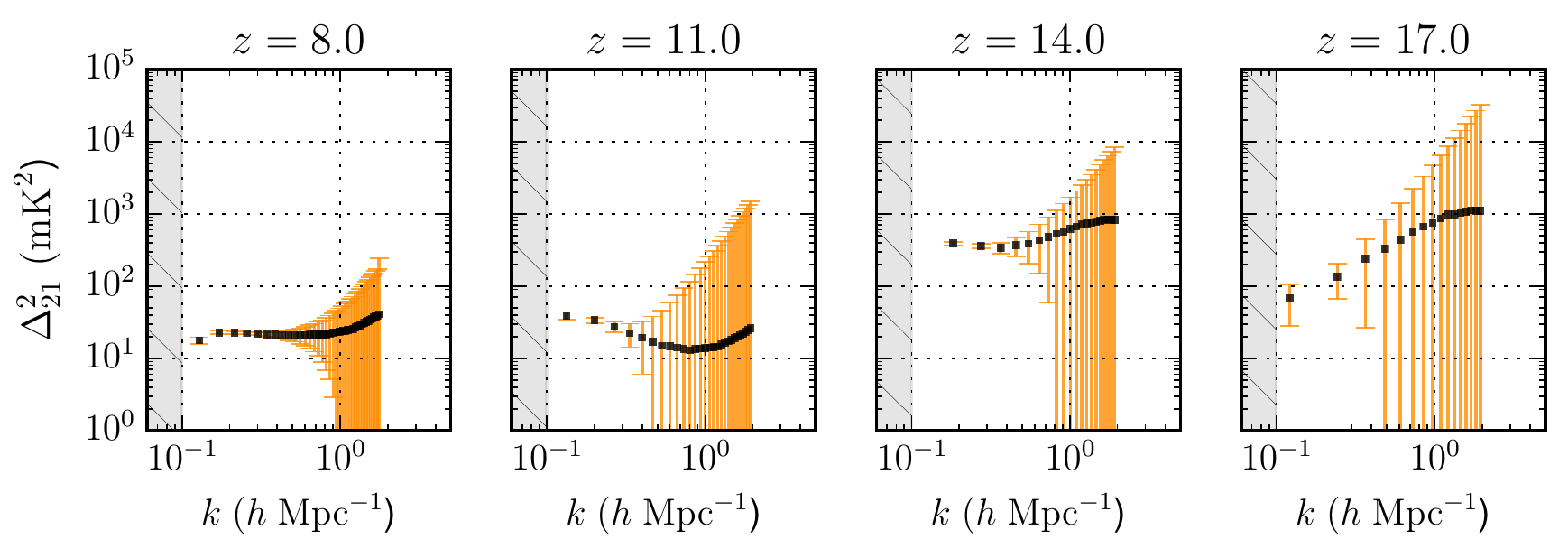}
\caption{A mock observation of the \tocm power spectrum created from an underlying ``truth'' realization of \texttt{21cmFAST} with error bars corresponding to the projected sensitivity of the HERA331 experiment after a single observing season. The grey-hatched region to the left denotes inaccessibility due to foreground dominated $k$ modes. Although we display only four redshifts, the entire mock observation contains the \tocm power spectrum from $5<z<25$ in steps of $\Delta z=0.5$.}
\end{figure*}

\subsubsection{Cosmological Parameters}
A previous study utilizing the Fisher Matrix approach found that even though cosmological parameters have precise constraints from other cosmological probes such as the \emph{Planck} satellite, their residual uncertainties introduce a non-negligible effect on the \tocm power spectrum and thus degrade the constraints one can place on astrophysical parameters using $21\,\textrm{cm}$ measurements \citep{Liu2016a}. Stated another way, by excluding cosmological parameters from a joint fit, we would be falsely \emph{overconstraining} the astrophysical parameters. Additionally, besides their degradation of astrophysical parameter constraints, we would also like to be able to constrain cosmology with the rich amount of information the \tocm signal provides us. We pick $\{\se, \Hn, \Ombhh, \Omchh, \ns\}$ as our cosmological parameter set. This particular parameterization is selected to match the current \texttt{21cmFAST} cosmological inputs and is done merely for convenience. It may be worth investigating in future work if other $\Lambda$CDM parameterizations are more suitable for \tocm constraints. In terms of \texttt{21cmFAST}, all of the chosen parameters play a role in setting the initial conditions for the density field, and $\Ombhh$, $\Omchh$ and $\Hn$ are furthermore directly related to the definition of the \tocm brightness temperature (\autoref{eqn:dTb}). Some of these cosmological parameters also play a role in transforming our observed coordinates on the sky into cosmological distance coordinates (see \autoref{eqn:noise}). While we do not include these effects into this study, a complete analysis would require such a consideration, which will be addressed in future work. Our fiducial values for the cosmological parameters are ($\se$, $h$, $\Ombhh$, $\Omchh$, $\ns$) = (0.8159, 0.6774, 0.0223, 0.1188, 0.9667), which are consistent with recent \emph{Planck} results \citep{Planck2016}. Because $\se$ and $H_{0}$ are not directly constrained by \emph{Planck} but are derived parameters in their $\Lambda$CDM parameterization, we use the \texttt{CAMB} code \citep{Lewis1999} to map the parameter degeneracies of $A_{s}$ (the normalization of the primordial perturbation power) and $\theta_{\textrm{MC}}$ (the \texttt{CosmoMC} code approximation for the angular size of the sound horizon at recombination; \citealt{Lewis2002}) onto that of $\se$ and $H_{0}$ respectively.


\section{Forecasted HERA331 Constraints}
\label{sec:forecast}

Here we forecast the ability of the HERA experiment to constrain the eleven parameter model described above. Before we present the parameter constraints, however, we must discuss how we construct our likelihood function and account for how errors in the emulator prediction can affect the final likelihood statistic. We begin by creating a mock observation for the HERA experiment and accounting for known systematics like bright foreground contamination.

\subsection{Interferometer Sensitivity Model}
\label{sec:sensitivity}
To create a mock \tocm power spectrum observation, we run \texttt{21cmFAST} with ``true'' model parameters set to the fiducial values described in \autoref{sec:parameterization}. The initial conditions of the density field are generated with a different random seed than what was used to construct the training set realizations.

Uncertainty in the \tocm power spectrum at the EoR comes from three main sources, (i) thermal noise of the instrument, (ii) uncertainty in foreground subtraction, and (iii) sampling (or cosmic) variance of our survey. For portions of Fourier space that are clean of foregrounds, the variance of the power spectrum at an individual $\bm{k}$ mode from the two remaining sources of uncertainty can be written as 
\begin{align}
\label{eqn:noise}
\sigma^{2}(\bm{k}) \approx \left[X^{2}Y\frac{k^{3}}{2\pi^{2}}\frac{\Omega^{\prime}}{2t}T^{2}_{\text{sys}} + \widehat{\Delta^{2}_{21}}(\bm{k})\right]^{2},
\end{align}
where the first term is the thermal noise ($k=|\bm{k}|$), and the second term is the sampling variance uncertainty at each individual $\bm{k}$ mode \citep{Pober2013a}. In the first term, $X^{2}Y$ are scalars converting angles on the sky and frequency spacings to transverse and longitudinal distances in $h\ \mathrm{Mpc^{-1}}$, and $\Omega^{\prime}$ is the ratio of the square of the solid angle of the primary beam divided by the solid angle of the square of the primary beam \citep{Parsons2014}. The total amount of integration time on a particular $k$ mode is $t$, and $T_{\text{sys}}$ is the system temperature taken to be the sum of a receiver temperature at 100 K and a sky temperature at $60\lambda^{2.55}$ K, where $\lambda$ has units of meters \citep{Parsons2014}. To compute the variance on the 1D power spectrum, $\mathrm{Var}[\DD(k)]$, from the above variances on the 2D power spectrum, we bin into annuli of constant scalar $k$ and add the variances reciprocally \citep{Pober2013a}.

We perform these calculations with the public Python package \texttt{21cmSense},\footnote{\url{https://github.com/jpober/21cmSense}} which takes as input a specification of the interferometer design and survey parameters \citep[see][]{Parsons2012b,Pober2014}. We assume a HERA-like instrument with a compact, hexagonal array of 331 dishes that each span 14-m in diameter \citep{Dillon2016, DeBoer2017}. We further assume the observations are conducted for 6 hours per day spanning a 180 day season for a total of 1080 observing hours. Within an instrumental bandpass spanning $50$-$250\,\textrm{MHz}$, we construct power spectra from $5 < z < 25$ in co-evolution redshift bins of $\Delta z = 0.5$. We also adopt the set of ``moderate'' foreground assumptions defined in \texttt{21cmSense}. This assumes that in a cylindrical Fourier space decomposed into wavenumbers perpendicular ($\kperp$) and parallel ($\kpara$) to the observational line-of-sight, the foreground contaminants are limited to a characteristic ``foreground wedge" at low $\kpara$ and high $\kperp$\citep[see e.g.,][]{Datta2010, Morales2012, Trott2012, Parsons2012a, Pober2013b, Liu2014a, Liu2014b}. One then pursues a strategy of foreground avoidance (rather than explicit subtraction) under the approximation that outside the foreground wedge there is a foreground-free ``EoR window". To be conservative, we impose an additional buffer above the foreground wedge of $k_{\parallel}=0.1\ h\ \mathrm{Mpc}^{-1}$ to control for foreground leakage due to inherent spectral structure of the foregrounds. The selection of this buffer is motivated by observations of \citet{Pober2013b}, who made empirical measurements of the foreground wedge as seen by the PAPER experiment at redshift $z\sim8.3$. For our sensitivity calculations, we impose a constant buffer at all redshifts, even though one would expect foreground wedge leakage to evolve with redshift just as the wedge itself evolves with redshift. Intuitively, we expect foreground leakage to have the same redshift dependence as the wedge: at higher redshifts foreground leakage reaches to higher $k_\parallel$ because the power spectrum window functions become more elongated along the $k_\parallel$ direction \citep[see][]{Liu2014a}. This means that our assumed buffer of $k_\parallel = 0.1\ h\ \mathrm{Mpc}^{-1}$ is over-conservative for $z < 8.3$ and under-conservative for $z > 8.3$.

We note that the sensitivity projections from \texttt{21cmSense} are assumed to be uncorrelated across $k$, meaning that our $\Sigma_{\mS}$ is diagonal. While this is not strictly true for a real experiment it is often an assumption made in parameter constraint forecast studies, with the reasoning that via careful binning in the $u-v$ plane informed by the extent of the telescope?s primary beam response, one can minimize the correlation between different $u-v$ modes. For more detailed discussions of foreground avoidance and subtraction techniques for \tocm interferometers, we defer the reader to \citet{Pober2014}. \autoref{fig:mock_obs} shows the resulting sensitivity projection produced by applying the above calculations to our truth \texttt{21cmFAST} realization.

\subsection{Constructing the Likelihood}
\label{sec:likelihood}
The likelihood function describes the ability of our observations to constrain the model parameters. This could in principle contain data from multiple observable probes of reionization. For this study, we focus solely on the likelihood function for the \tocm power spectrum, but future work will investigate extending this formalism to incorporate other EoR probes. Our \tocm power spectrum likelihood function can be written up to a constant as
\begin{align}
\label{eqn:loglike}
\ln{\mcL}(\by|\btheta) &\propto -\frac{1}{2}(\by-\bd)^{T}\bSigma_{\mS}^{-1}(\by-\bd),
\end{align}
where $\bSigma_{\mS}$ is a diagonal covariance matrix containing the observational survey error bars (including both thermal noise and sample variance as described in \autoref{sec:sensitivity}), $\by$ are our observations of the \tocm power spectrum spanning a redshift range of $5<z<25$ and scale range $0.1 < k < 2\ h\ \mathrm{Mpc^{-1}}$, and $\bd$ are the model predictions evaluated at some point in parameter space $\btheta$. In \citet{Greig2015a}, who similarly investigated parameter constraints with \texttt{21cmFAST}, the authors add an additional $20\%$ error on the sampled \tocm power spectra along the diagonal of their likelihood covariance matrix to account for the $\sim10$'s of percent difference in the power spectra between \texttt{21cmFAST} and detailed numerical simulations \citep{Mesinger2011, Zahn2011}. In this work, we do not include this term because we do not claim that our constraints with \texttt{21cmFAST} are representative of the constraints that a numerical simulation might place. Rather, because we are using \texttt{21cmFAST} as our model, we have implicitly performed model selection up-front, and can only place quantifiable constraints on the parameters of this model. The $\bSigma_{\mS}$ term in our likelihood function therefore contains only the variance of the uncorrelated survey sensitivity (calculated in \autoref{sec:sensitivity}) along its diagonal.

If we were directly using our simulation to generate our model vectors $\bd$, then this is indeed the likelihood function we would seek to minimize to produce our parameter constraints. However, in the regime where the simulation is either too computationally expensive or too slow to directly evaluate, we replace it with its emulated prediction $\bd_{\mE}$. The emulated prediction is an approximation, and can be related to the true simulation output as
\begin{align}
\label{eqn:delta}
\bd = \bd_{\mE} - \bdelta,
\end{align}
where $\bdelta$ is offset of the emulator from the simulation. Naturally, we do not know $\bdelta$ any more than we know $\bd$ without directly evaluating the simulation; however, treating $\bdelta$ as a random variable, we do have an estimate of its probability distribution given to us by either our Gaussian Process fit or by cross validation.

If we treat $\bdelta$ as a Gaussian random variable, then its probability distribution is given as
\begin{align}
\label{eqn:gauss_dist}
\bdelta \sim \mathcal{N}(0,\bSigma_{\mE}),
\end{align}
where $\bSigma_{\mE}$ is the covariance matrix describing the uncertainty on $\bd_{\mE}$, and in principle can contain both uncorrelated errors along its diagonal terms and correlated errors in its off-diagonal terms. This variable can be estimated by using the output of the Gaussian Process, or it can be empirically calculated from cross validation. In the context of our worked example, we will always defer to calculating this variable empirically via cross validation, which means that $\bSigma_{\mE}$ is constant throughout the parameter space.

We can, and should, account for the fact that errors in our emulator's model predictions will induce errors into our likelihood. Writing out the likelihood in terms of \autoref{eqn:delta}, we see that
\begin{align}
\label{eqn:lnlike_start}
\ln{\mcL}(\by|\btheta) &\propto -\frac{1}{2}(\by-\bd_{\mE}+\bdelta)^{T}\bSigma_{\mS}^{-1}(\by-\bd_{\mE}+\bdelta).
\end{align}
If we do not know the precise value of $\bdelta$, we can propagate its uncertainty into the likelihood by marginalizing over its possible values. The derivation is given in \autoref{sec:gauss_int}, and the result is that \autoref{eqn:lnlike_start} can be cast as
\begin{align}
\label{eqn:final_lnlike}
\ln{\mcL}(\by|\btheta) &\propto -\frac{1}{2}(\by-\bd_{\mE})^{T}\left(\bSigma_{\mS}+\bSigma_{\mE}\right)^{-1}(\by-\bd_{\mE}),
\end{align}
where the marginalization process has left with a larger effective covariance matrix that is the direct sum of the error matrices of our two sources of error (observational error and emulator error). The inflated covariance matrix means we will recover broader constraints than if we had not included the emulator error budget. This is actually a desirable trait; it is better to recover broader constraints and be more confident that the truth parameters fall within those constraints rather than bias ourselves into falsely over-constraining regions of parameter space. In \autoref{sec:discussion}, we provide a test case for our emulator to see if it can indeed inform us when there has been a failure, such as the training set missing the location of some of the ``true'' parameters of a mock observation.

The likelihood function defined above will be what we use to directly constrain our model parameters of-interest. Because this function is non-analytic, we use a Markov-Chain Monte Carlo (MCMC) sampling algorithm to find this function's peak and characterize its topology. There are a number of samplers that are well suited for this task. Our emulator employs \texttt{emcee}, the ensemble sampler algorithm from \citet{Foreman-Mackey2013}, which is itself based on the affine-invariant sampling algorithm described in \citet{Goodman2010}.

\begin{figure*}
\label{fig:recplot_initial}
\centering
\includegraphics[scale=0.75]{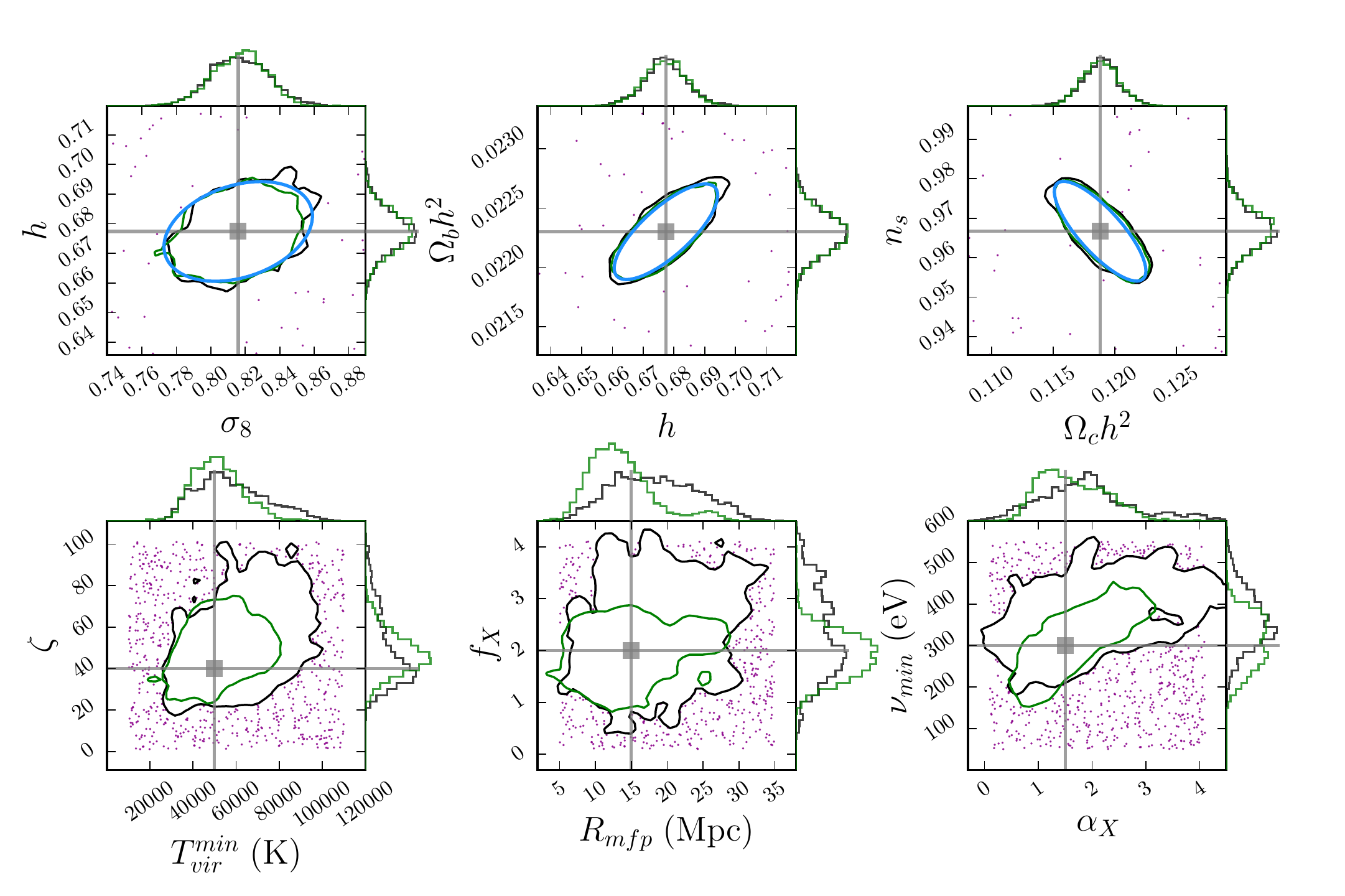}
\caption{Posterior constraints for our initial parameter space exploration. The black contours represent 95\% posterior credibility after emulating over our rectangular LH-design training set (shown as purple points). The green contours represent 95\% posterior credibility after emulating over the LH training set plus a second, spherical training set populated within the contours of the initial constraints. The blue ellipses over the cosmological parameters show the 95\% probability contour of our \emph{Planck} prior distribution. The grey square shows the true underlying parameters of the observation. The histograms adjacent to the contour plots show the marginalized posterior distribution across each model parameter.}
\end{figure*}

\subsection{Broad Parameter Space Search}
\label{sec:param_explore}

To produce parameter constraints with an emulator, we must first construct a training set spanning the regions of parameter space we would like our MCMC sampler to explore. Due to the finite size of any training set, we need to set hard limits \emph{a priori} on the breadth of the training set in parameter space. Our prior distribution on the model parameters is a straightforward way to make this choice. The astrophysical parameters of EoR and EoH, however, are highly unconstrained and in some cases span multiple orders of magnitude. In order to fully explore this vast parameter space with the emulator, we are left with a few options: (i) we could construct a sparse and wide training set, emulate at a highly approximate level, MCMC for the posterior distribution and then repopulate the parameter space with more training samples in the region of high probability and repeat, or (ii) use a gradient descent method to locate the general location of maximum probability. Both require direct evaluations of the simulation, but the former can be done in batch runs on a cluster and the latter is a sequential, iterative process (although it is typically not as computationally demanding as a full MCMC). For this work, we choose the former, and construct a wide rectangular training set from a Latin-Hypercube design consisting of $15\times10^{3}$ points. For one parameter in particular, $\fX$, we do not cover the entire range of its currently allowed values. In order to exhaustively explore the prior range of $\fX$, one might consider performing an initial gradient descent technique to localize its value. Because gradient descent algorithms are common in the scientific literature, we do not perform this test and assume we have already localized the value of $\fX$ to within the extent of our initial training set, or assume we are comfortable adopting a prior on $\fX$ spanning the width of our initial training set.

\begin{figure*}
\label{fig:triplot}
\centering
\includegraphics[scale=0.28]{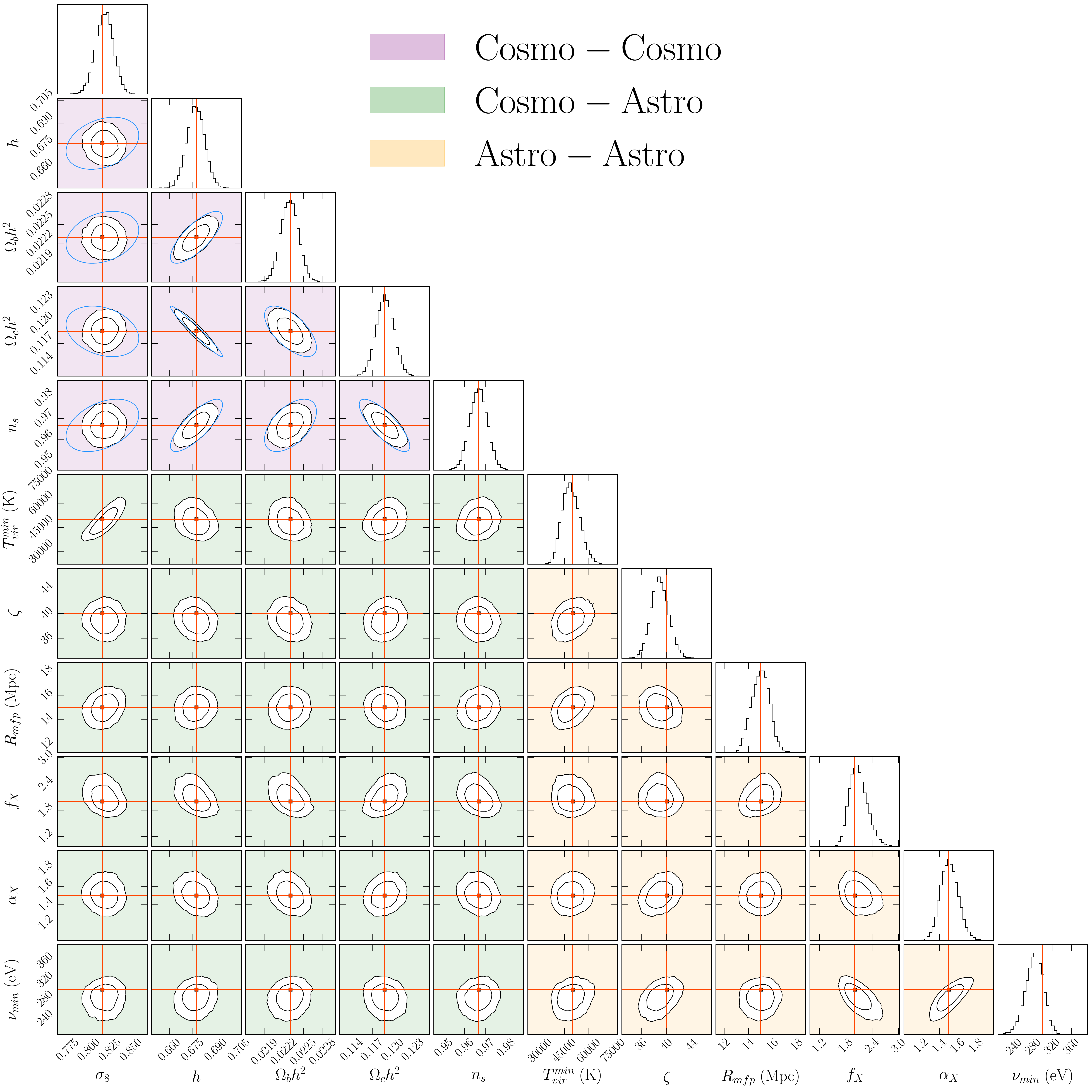}
\caption{The joint posterior distribution of the eleven-parameter model, showing the 68\% and 95\% credible regions of the pairwise covariances (off-diagonal) and their marginalized distribution across each model parameter (diagonal). Purple-shaded boxes represent pairwise covariances between cosmological parameters; green-shaded boxes represent cosmological-astrophysical covariances, and yellow-shaded boxes represent astrophysical covariances. Blue contours on the cosmological covariances indicate the 95\% credible region of the adopted prior distribution consistent with \emph{Planck}. The underlying true parameters of the observation are marked as red squares with crosshairs.}
\end{figure*}

We use this initial training set to solve for an estimate of the hyperparameters for our Gaussian Process kernel as detailed in \autoref{sec:hyper_params}. With a training set consisting of over $10^4$ points, we do not solve for a global predictive function of the eigenmode weights, but use a variant of the Learn-As-You-Go algorithm described in \autoref{sec:GPR} for emulation. We $k$-fold cross validate on the training set and find that we can emulate the power spectra to accuracies ranging in the 50\%-100\% level depending on the redshift and $k$ mode. While this is by no means ``high-precision'' emulation (and will pale in comparison to the precision achieved in our final emulator runs for producing the ultimate parameter constraints), it is enough to refine our rough estimate of the location of the MAP point. We incorporate these projected emulator errors into our likelihood as described in \autoref{sec:likelihood}. We adopt flat priors over the astrophysical parameters and covarying priors on the cosmological parameters representative of the \emph{Planck} base TT+TE+EE+low-$\ell$ constraint, whose covariance matrix can be found in the \texttt{CosmoMC} code \citep{Lewis2002}.

We show the results of our initial parameter space search in \autoref{fig:recplot_initial}, where the black contours represent the 95\% credible region of our constraint and the histograms show the posterior distribution marginalized across all other model parameters. The purple points in \autoref{fig:recplot_initial} show samples from the initial LH training set, demarcating its hard bounds. The blue contours on the cosmological parameters show the 95\% credible region of the prior distribution, showing that at this level of emulator precision the posterior distribution across the cosmology is dominated by the strong \emph{Planck} prior. Even while emulating to a highly approximate level, we find that we can recover a rough and unbiased localization of the underlying truth parameter values. After localization, we can choose to further refine the density of our training set to produce better estimates of the MAP point and ensure we are converging upon the underlying true parameters. To do this, we extend the training set with an extra 5,000 samples sampled from a spherical multivariate Gaussian located near the truth parameters with a size similar to the width of the posterior distribution. The 95\% credible region of the parameter constraints produced using an updated training set of 20,000 samples is shown in \autoref{fig:recplot_initial} as the green contours, which shows an improvement in the MAP localization.

\subsection{Posterior Characterization}
\label{sec:constraints}

With a reasonable estimate of the MAP location in hand, we now construct a dense training set so that we may emulate to higher precision. To do this, we build another training set consisting of 5000 samples from a packed Gaussian distribution and 500 samples from a LHSFS design (see \autoref{sec:exp_design}) with a location near the truth parameters and size similar to the posterior distribution found in \autoref{sec:param_explore}. To assess the accuracy of the emulator trained over this training set, we $5$-fold cross validate over a subset of the data in the core of the training set. The results can be found in \autoref{fig:cv_plots}, which shows we can emulate the power spectra to an accuracy of $\sim10\%$ for most of the data. More importantly, however, \autoref{fig:cv_plots} shows that the emulator error is always lower than the inherent observational survey error, and for the majority of the data is considerably lower. Nonetheless, we account for these projected emulator errors by adding them in quadrature with the survey error bars as described in \autoref{sec:cross_valid}. Our MCMC run setup involves 300 chains each run for $\sim$5,000 steps, yielding over $10^{6}$ posterior samples. On a MacPro Desktop computer, this entire calculation takes $\sim12$ hours and utilizes $\sim10$ GB of memory.

The final characterization of the posterior distribution is found in \autoref{fig:triplot}, where we show its marginalized pairwise covariances between all eleven model parameters and its fully marginalized distributions along the diagonal. With the exception of $\se$, the cosmological constraints are mostly a reflection of the strong \emph{Planck} prior distribution (shown as blue contours). Compared to previous EoR forecasts of \citet{Pober2014, Ewall-Wice2016a, Greig2016}, the strength of the EoR parameter degeneracies are weakened due to the inclusion of cosmological physics that washes out part of the covariance structure. This importance is exemplified by the strong degeneracy between the amplitude of clustering, $\se$, and the minimum virial temperature, $\Tvirmin$. At a particular redshift, an increase in $\se$ increases the number of collapsed dark matter halos. At the same time, an increase in $\Tvirmin$ suppresses the number of collapsed halos that can form stars, meaning they balance each other out in terms of their effect on the number of star forming halos present at any particular redshift. This degeneracy on the overall timing of EoR between these parameters is clearly seen in the animation tied to \autoref{fig:21cmFAST} (see caption). 

\begin{figure}
\label{fig:marghist}
\centering
\includegraphics[scale=0.55]{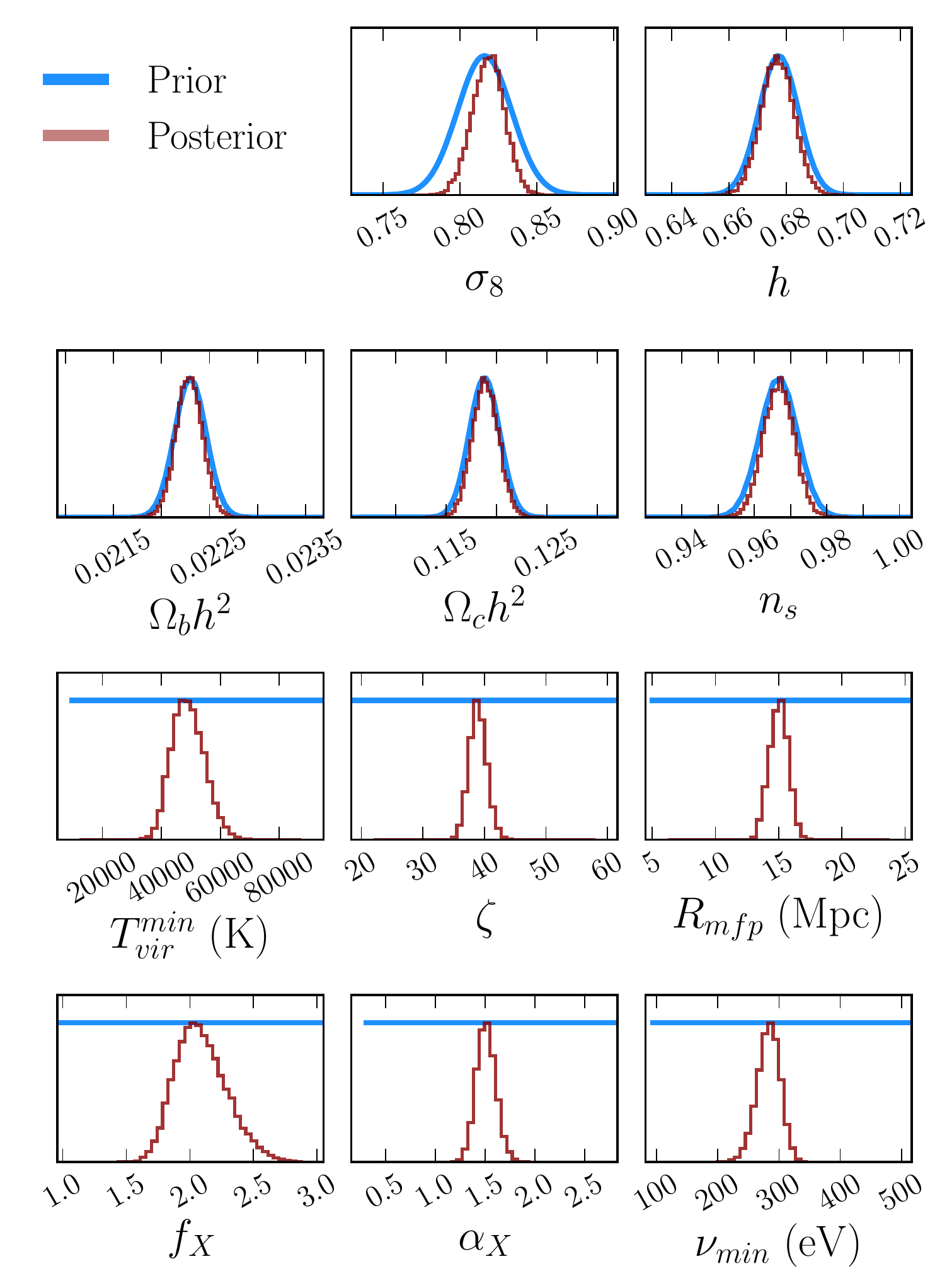}
\caption{The posterior distribution of \autoref{fig:triplot} for each model parameter marginalized across all other parameters, compared against the adopted prior distributions. We adopt priors on the cosmological parameters consistent with \emph{Planck} constraints, and adopt flat priors across the astrophysical parameters. We find that HERA will be able to produce $\sim10\%$ level constraints on the astrophysical parameters and will help strengthen constraints on $\se$.}
\end{figure}

Compared to the recent work of \citet{Greig2017b}, who performed a full MCMC over EoR and EoH parameters with \texttt{21cmFAST} assuming a HERA-331 experiment, our constraints are slightly stronger. This could be for a couple of reasons, including (i) the fact that they add an additional 20\% modeling error onto their sampled power spectra and (ii) their choice of utilizing power spectra across 8 redshifts when fitting the mock observation, compared to our utilization of power spectra across 37 different redshifts when fitting to our mock observation.

The posterior distributions for each parameter marginalized across all others are shown in \autoref{fig:marghist}, where they are compared against their input prior distributions. We see that the HERA331 experiment, with a moderate foreground avoidance scheme, will nominally place strong constraints on the EoR and EoH parameters of \texttt{21cmFAST} with respect to our currently limited prior information. For the cosmological parameters, the HERA likelihood alone is considerably weaker than the \emph{Planck} prior; however, we can see that a HERA likelihood combined with a \emph{Planck} prior can help strengthen constraints on certain cosmological parameters. Because \tocm experiments are particularly sensitive to the location of the redshift peaks of the \tocm signal,\footnote{Strong peaks and dips in the $z$ evolution of $\DD$ mean that slight deviations along $z$ produce large deviations in $\DD$.} parameters like $\se$, which control the overall clustering and thus affect the timing of reionization, are more easily constrained. Going further, \citet{Liu2016a} showed that one can produce improved CMB cosmological parameter constraints by using \tocm data to constrain the prior range of $\tau$, which is a CMB nuisance parameter that is strongly degenerate with $\se$ and thus degrades its constraining power. Our \tocm power spectrum constraint on $\se$ shown above does not include this additional improvement one can achieve by jointly fitting \tocm and CMB data, which is currently being explored.


\section{Discussion}
\label{sec:discussion}
Here we discuss performance tests that help to further validate the efficacy of the emulator algorithm. We address the issue of what happens when the underlying true parameters lie at the edges or outside of the hard bounds of our training set, and make a direct comparison of the constraints produced by our emulator and a traditional brute-force MCMC algorithm.

\begin{figure}
\label{fig:directMCMC}
\centering
\includegraphics[scale=0.4]{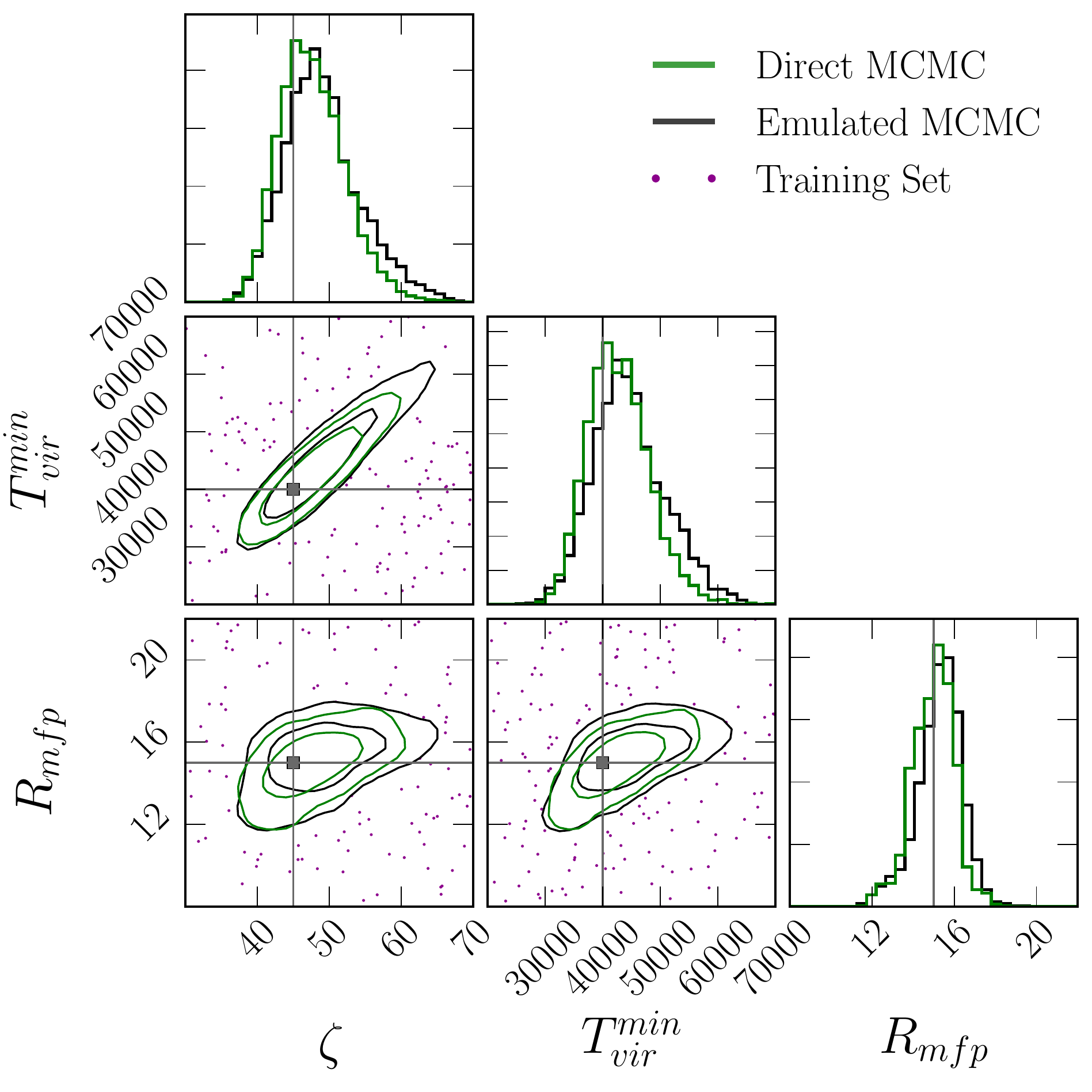}
\caption{Emulator performance test comparing the constraints from the emulator (black) against brute-force constraints which directly evaluate the simulation (green). Both are able to produce unbiased constraints on the underlying ``truth'' parameters of the mock observation (square). The training set samples used to construct the emulator are shown in the background (purple points).}
\end{figure}

\begin{figure*}
\label{fig:recplot_outofbounds}
\centering
\includegraphics[scale=0.65]{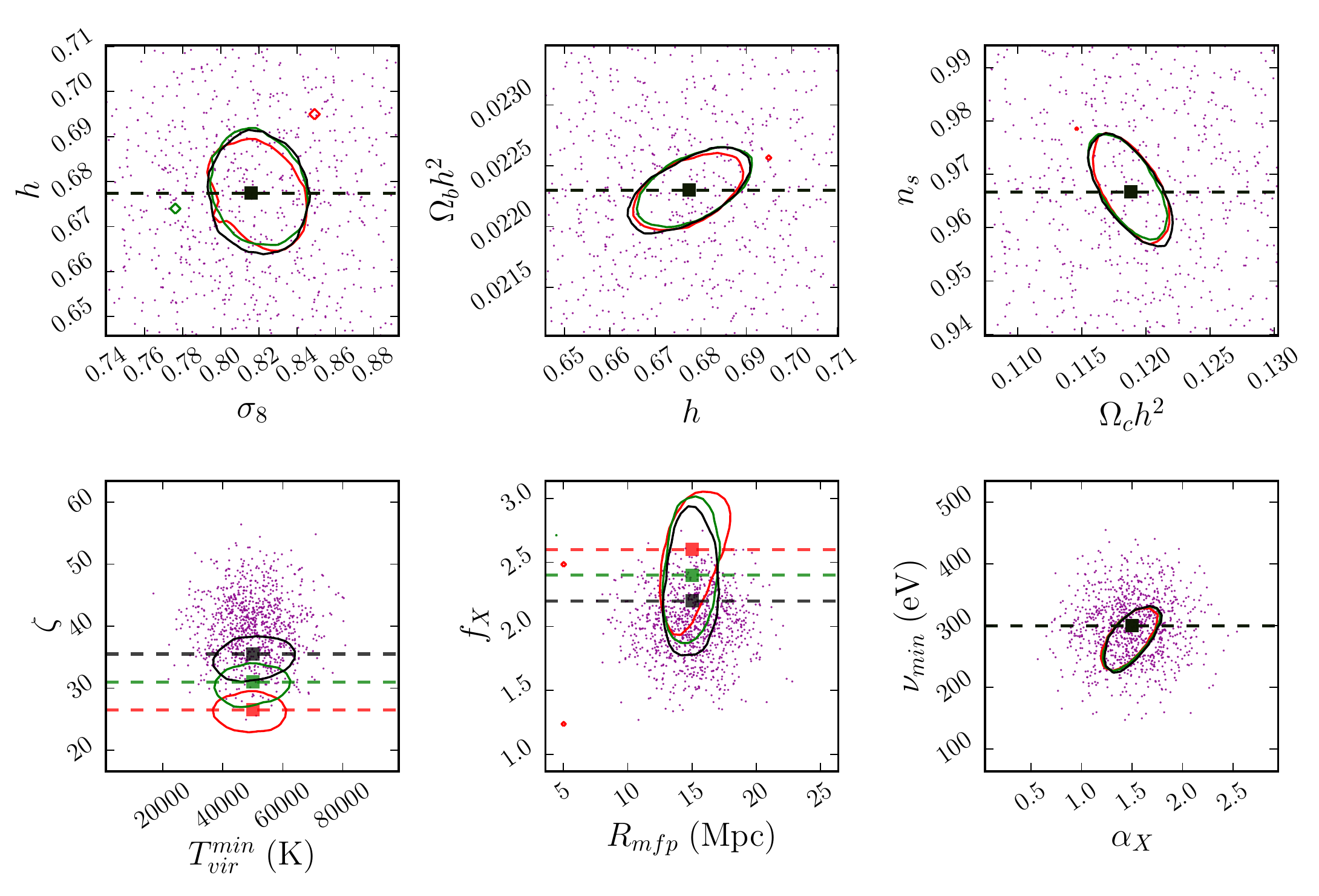}
\caption{95\% credible regions of the posterior distribution while moving the true parameters of the mock observation away from the center of the training set, demonstrating the ability of the emulator to recover unbiased MAP constraints even when the training set does not directly overlap with the underlying ``truth'' parameters.}
\end{figure*}

\subsection{Comparison Against Direct MCMC}
\label{sec:directMCMC}

An important performance test of the emulator algorithm is to compare its parameter constraints against the constraints produced by brute-force, where we directly call the simulation in our MCMC sampler. Of course we cannot do this for the simulation we would like to use---hence the need for the emulator---but we can do this if we use a smaller and faster simulation. For this test, we still use \texttt{21cmFAST} but only generate the power spectra at $z=\{8,9,10\}$ and ignore the spin temperature contribution to $\DD$, which drastically speeds up the simulation. In addition, we use a smaller simulation box-size and use a coarser resolution which yields additional speed-ups. We also restrict ourselves to varying only the three EoR astrophysics parameters described in \autoref{sec:eor_params}, meaning we achieve faster MCMC convergence. Using a coarser resolution and ignoring spin temperature fluctuations means the simulation is less accurate, but for the purposes of this test the simulation accuracy is irrelevant: we merely want to gauge if the emulator induces significant bias into constraints that we would otherwise produce by directly using the simulation. 

Our mock observation is constructed using a realization of \texttt{21cmFAST} with fiducial EoR parameters ($\zeta$, $\Tvirmin$, $\Rmfp$) = ($45$, $4\times10^{3}$ K, $15$ Mpc), and with the same fiducial cosmological parameters of \autoref{sec:constraints}. We place error bars over the fiducial realization using the same prescription of that described in \autoref{sec:sensitivity}, corresponding to the nominal sensitivity projections for the HERA331 experiment under ``moderate'' foreground avoidance. Similar to \autoref{sec:sensitivity}, we fit the power spectra across $0.1 \le k \le 2\ h\ \mathrm{ Mpc^{-1}}$ in our MCMC likelihood function calls.

The result of the test is shown in \autoref{fig:directMCMC}, where we plot the emulator and brute-force posterior constraints, as well as the training set samples used to construct the emulator. We find that the emulator constraints are in excellent agreement with the constraints achieved by brute-force. In the case where the emulator constraints slightly deviate from the brute-force constraints (in this case high $\zeta$ and high $\Tvirmin$), the emulator deviations are conservative relative to the brute-force contours. In other words, the emulator constraints are always equal to or broader than the brute-force constraints, and do not falsely over-constrain the parameter space or induce systematic bias into the recovered MAP.

\subsection{Training Set Miscentering}
\label{sec:offset_recovery}
The ability of the emulator to produce reliable parameter constraints hinges principally on the assumption that the true parameter values lie within the bounds of the training set. If this is not the case, the emulator cannot make accurate predictions of the simulation behavior and is making a best guess based on extrapolation. In the case that emulator errors are not accounted for, this can lead to artificial truncation of the posterior distribution and create a false, over-constraining of the parameter space. This was observed to be problematic for a small number of figures in the 2015 \emph{Planck} papers. Though the underlying cosmological constraints were unaffected, some illustrative plots employed an emulator-based method that seemed to be in tension with a more accurate direct MCMC method because the underlying parameters lay outside of the emulator's training set \citep{Addison2016}. It is therefore crucial to be able to assess if our training set encompasses the underlying truth parameters or if the training set has been miscentered. If the emulator can alert us when this is the case, we can repopulate a new training set in a different location and have greater confidence that the emulator is not falsely constraining the parameter space due to the finite width of the training set.

Given our method in \autoref{sec:forecast} for localizing the parameter space via a sequence of training sets that iteratively converge upon the general location of the underlying true parameters, it is natural to ask, what if we made our final, compact training set a little too compact and missed the underlying MAP? How can we assess if this is the case, and if so, where do we populate the new training set? The most straightforward answer is to look at the posterior constraints compared to the width of the training set: if the posterior constraints run-up against the edge of the training set significantly, this may be an indication that we need to move the training set in that direction.

We perform such a test using our final compact training set and shift the position of mock observation's underlying truth parameters to the edges of the training set for parameters $\zeta$ and $\fX$: two particularly unconstrained parameters. \autoref{fig:recplot_outofbounds} shows the result, demonstrating the emulator's ability to shift the posterior contours when it senses that the MAP lies at the edge of the training set. In this case, we would know to generate more training samples near the region of high probability and retrain our emulator.

\section{Conclusions}
\label{sec:conclusion}

The next generation of \tocm radio interferometric experiments with raw sensitivities at fiducial EoR levels are currently being built. The next few years will likely see either a detection and characterization of the \tocm power spectrum or strong upper limits. However, interpreting these observations and connecting them to the high dimensional and weakly constrained parameter space of Cosmic Dawn is not straightforward. This is in part because the relevant physics spans many order of magnitude in dynamic range and is non-linear, making detailed simulations computationally expensive, slow, and not conducive for iteration within an MCMC framework. While semi-numerical approaches have made progress exploring this parameter space, even they can have difficulty achieving speeds quick enough for MCMC techniques.

One way to address this challenge is to build an emulator for the simulation, which mimics the simulation output and is generally many orders of magnitude faster to evaluate. Emulators yield a few crucial advantages for parameter constraints over a direct MCMC approach. First, after the overhead of building the training set and training the emulator, running a parameter constraint analysis with an emulator is extremely cheap and quick. This feature is beneficial for MCMC repeatability: if we change our instrumental or survey covariance matrix, add more data to our observations, discover a bug in our data analysis pipeline, or find that a particular MCMC sampler is not exploring the parameter space properly, we would need to completely re-run these chains. Without an emulator, this could become a computationally prohibitive cost even for the most optimized semi-numerical simulations.

However, emulators also have their challenges. Most importantly, emulators are dependent on a training set, which will invariably have a finite size. This means we must a priori select a finite range over which our emulator is valid. This choice can be particularly hard to make for parameters that are highly unconstrained. This can be overcome by prefacing emulation with a gradient descent algorithm to localize parameters that are particularly unconstrained, or by incorporating prior information on these parameters from other probes.

In preparation for analyzing data sets from current and upcoming \tocm experiments, we have built a fast emulator that can mimic simulations of Cosmic Dawn to high precision. We review the emulator algorithm present in our publicly available \texttt{emupy} and \texttt{pycape} packages, and discuss techniques for data compression, Gaussian Process regression and cross validation. We then apply our emulator to a simulation of Cosmic Dawn and demonstrate its ability to take a mock observation of the \tocm signal and produce constraints on fundamental astrophysical and cosmological parameters. We find that a characterization of the \tocm power spectrum from the upcoming Hydrogen Epoch of Reionization Array (HERA) will considerably narrow the allowed parameter space of reionization and X-ray heating parameters, and could help strengthen the constraints on $\se$ already set by \emph{Planck}. The forecast presented in this work used an MCMC setup with 300 chains, each run for 5,000 steps, which took $\sim12$ hours on a MacPro desktop. While this forecast utilized a specific simulation of Cosmic Dawn, the emulator package and the parameter constraint framework outlined in this work are entirely independent: we could in principle emulate a whole suite Cosmic Dawn simulations ranging in their numerical implementations with only minor changes to the procedure outlined in this work. Although in this work we focus solely on the constraining power of the \tocm power spectrum, the emulator framework can also be used to incorporate information from other probes of reionization, such as the hydrogen neutral fraction, averaged brightness temperature, electron scattering optical depth, galaxy clustering statistics and higher order statistics probes of the \tocm field. Future work synthesizing these observables under the emulator framework will address this, enabling \tocm intensity mapping efforts to live up to their theoretical promise of constraining a wide breadth of astrophysical and cosmological physics.

\acknowledgments
The authors would like to thank Grigor Aslanyan, Michael Betancourt, Josh Dillon, Danny Goldstein, Raul Monsalve, Danny Jacobs, Uro\v{s} Seljak, and Martin White  for helpful discussions. AM and BG acknowledge funding from the European Research Council (ERC) under the European Union's Horizon 2020 research and innovation program (Grant Agreement No. 638809: AIDA). AL, ARP, and NSK acknowledge support from the University of California Office of the President Multicampus Research Programs and Initiatives through award MR-15-328388, as well as from NSF CAREER award No. 1352519, NSF AST grant No.1129258, NSF AST grant No. 1410719, and NSF AST grant No. 1440343. AL acknowledges support for this work by NASA through Hubble Fellowship grant \#HST-HF2-51363.001-A awarded by the Space Telescope Science Institute, which is operated by the Association of Universities for Research in Astronomy, Inc., for NASA, under contract NAS5-26555. This research used resources of the National Energy Research Scientific Computing Center, a DOE Office of Science User Facility supported by the Office of Science of the U.S. Department of Energy under Contract No. DE-AC02-05CH11231.

\software{
The following publicly-available software was utilized in our work: 
21cmFAST \citep[v1.2,][]{Mesinger2011},
21cmSense \citep{Pober2013b,Pober2013a},
Astropy \citep{Astropy2013},
emcee \citep{Foreman-Mackey2013},
emupy (\url{https://doi.org/10.5281/zenodo.886043}),
corner.py (\url{https://doi.org/10.5281/zenodo.53155}),
CosmoMC \citep{Lewis2002},
Matplotlib (\url{https://doi.org/10.5281/zenodo.573577}),
pycape (\url{https://doi.org/10.5281/zenodo.886026}),
SciKit-Learn \citep{Pedregosa2012},
Scipy (\url{https://doi.org/10.1109/MCSE.2007.58}).
}

\appendix

\section{Emulator Error Propagation}
\label{sec:gauss_int}

In this Appendix, we derive \autoref{eqn:final_lnlike} from \autoref{sec:likelihood} for propagating emulator interpolation error into our final likelihood function. We start by assuming the emulator prediction $\bd_{\mE}$ is offset from the true simulation output $\bd$ by some amount $\bdelta$, such that $\bd \equiv \bd_{\mE} - \bdelta$. In practice we do not know $\bdelta$ precisely, but we do have an estimate of its possible values, given to us by our Gaussian Process fit (see \autoref{eqn:data_recon_cov}). In particular, we have an estimate of its probability distribution, modeled as a zero-mean Gaussian distribution with covariance given by $\bSigma_{\mE}$ (\autoref{eqn:gauss_dist}), which will act as our prior distribution on $\bdelta$.

Our likelihood function $\mcL$ is given by
\begin{align*}
\ln{\mcL}(\by|\btheta) &\propto -\frac{1}{2}(\by-\bd_{\mE}+\bdelta)^{T}\bSigma_{\mS}^{-1}(\by-\bd_{\mE}+\bdelta),
\end{align*}
where $\by$ is a vector containing the observations, $\bSigma_{\mS}$ a matrix containing the survey error bars, and both $\bd$ and $\bdelta$ are functions of the model parameters $\btheta$. Multiplying this likelihood with a Gaussian prior on $\bdelta$ yields the posterior distribution $\mathcal{P}$ given by
\begin{align*}
\ln{\mathcal{P}} \propto &-\frac{1}{2}(\by-\bd_{\mE}+\bdelta)^{T}\bSigma_{\mS}^{-1}(\by-\bd_{\mE}+\bdelta) - \frac{1}{2}\bdelta^{T}\bSigma_{\mE}^{-1}\bdelta\\
= &-\frac{1}{2}(\by-\bd_{\mE})^{T}\bSigma_{\mS}^{-1}(\by-\bd_{\mE})\\
&-\frac{1}{2}\bdelta^{T}\left(\bSigma_{\mS}^{-1}+\bSigma_{\mE}^{-1}\right)\bdelta + \left(\by-\bd_{\mE}\right)^{T}\bSigma_{\mS}^{-1}\bdelta
\end{align*}
where in the second expression we factored out a term that is independent of $\bdelta$.

We can account for $\bdelta$'s influence on $\mcL$ by marginalizing (i.e. integrating) over it. To do so, we make use of the identity 
\begin{align}
\label{eqn:gauss_int}
\int\exp\left[-\frac{1}{2}\bx^{T}\bA\bx+\bb\cdot\bx\right]d^{n}\bx = \sqrt{\frac{(2\pi)^{n}}{\det\bA}}\exp\left[\frac{1}{2}\bb^{T}\bA^{-1}\bb\right],
\end{align}
where $\bA$ is an $n\times n$ real, symmetric matrix, and $\bb$ and $\bx$ are both vectors of length $n$.
The resulting posterior distribution becomes
\begin{align*}
\ln{\mathcal{P}} \propto &-\frac{1}{2}\left(\by - \bd_{\mE}\right)^{T}\times\\
&\left[\bSigma_{\mS}^{-1}-\bSigma_{\mS}^{-1}\left(\bSigma_{\mS}^{-1}+\bSigma_{\mE}^{-1}\right)^{-1}\bSigma_{\mS}^{-1}\right]\left(\by-\bd_{\mE}\right),
\end{align*}
which can be simplified using identity
\begin{align*}
\left(\bA + \bB\right)^{-1} \equiv \bA^{-1} + \bA^{-1}\left(\bA^{-1}+\bB^{-1}\right)^{-1}\bA^{-1}
\end{align*}
to give
\begin{align}
\ln{\mathcal{P}} \propto -\frac{1}{2}\left(\by - \bd_{\mE}\right)^{T}\left(\bSigma_{\mS}+\bSigma_{\mE}\right)^{-1}\left(\by-\bd_{\mE}\right).
\end{align}
In other words, the emulator error covariance and the observational (or survey) error covariance simply add to form a new effective covariance that allows $\mcL$ to account for emulator error fluctuations in $\bd_{\mE}$. This result can also be reached by expressing $(\by-\bd_{\mE})$ as the sum of random variables $(\by-\bd)$ and $\bdelta$, which we can think of as the convolution of two Gaussian distributions. If we assume each are normally distributed random variables with covariance $\bSigma_{\mS}$ and $\bSigma_{\mE}$ respectively, then the probability distribution of their sum is equivalent to the convolution of their individual probability distributions. The convolution theorem then tells us that the variance of the normal distribution describing their sum is just the sum of their individual variances, or $\bSigma = \bSigma_{\mS}+\bSigma_{\mE}$.

\bibliographystyle{apj}
\bibliography{EoR}

\end{document}